\documentclass[aps,prd,twocolumn,nofootinbib]{revtex4-2}
\usepackage{graphicx}
\usepackage{amsmath,amssymb}
\usepackage[many]{tcolorbox}
\usepackage{comment}

\begin{document}

\title{
High-Energy Neutrinos from Black Hole Evaporation in Neutron Stars}
\author{Ioannis Dalianis}~
\email{ntalianis.ioannis@ucy.ac.cy}
\affiliation{Department of Physics, University of Cyprus, Nicosia 1678, Cyprus}


\begin{abstract}
We investigate the production of high-energy neutrinos from
microscopic black holes formed through the gravitational collapse of
asymmetric dark matter accumulated inside neutron stars. When Hawking
evaporation dominates over accretion, long-lived, feebly interacting
particles beyond the Standard Model escape the neutron star and
subsequently decay into high-energy neutrinos. We analyze the repeated
cycle of dark matter capture, black hole formation, and evaporation,
identifying two distinct regimes determined by the competition between
the dark matter thermalization time and the collapse cycle. In
particular, we identify a partially thermalized regime in which the
dark matter cloud evolves toward a quasi-stationary state with a
temperature significantly exceeding that of the neutron star core. We
derive the time-integrated Hawking emission, the resulting secondary
neutrino spectra, and the expected Galactic and diffuse extragalactic
neutrino fluxes. The predicted signal exhibits two distinctive signatures: a broad
neutrino spectrum with a characteristic energy scale set by the
initial Hawking temperature of the evaporating black hole, whose
spectral peak naturally lies above
$\mathcal{O}(10)$ TeV, and an extended Galactic component
strongly concentrated toward the Galactic Center.  Although the predicted event rates are generally small, the resulting
signal may contribute at the percent level to the observed Galactic
high-energy neutrino flux under favorable microscopic and
astrophysical conditions. The
proposed mechanism provides a new observational window on Hawking
evaporation through microscopic black holes continuously produced
inside neutron stars, linking dark matter, compact objects, black hole thermodynamics and
high-energy neutrino astronomy.
\end{abstract}

\maketitle

\section{Introduction}

Black holes occupy a unique position at the interface of gravitation and 
quantum theory. 
They are now routinely observed
over a wide range of masses,  from stellar-mass black holes detected
through gravitational waves \cite{KAGRA:2021vkt} to supermassive black holes imaged by the Event Horizon Telescope \cite{EHT:2019}, while compelling
evidence for intermediate-mass black holes has also emerged \cite{Greene:2019vlv}.
These observations have transformed black holes from theoretical
concepts into precision astrophysical laboratories. Gravitational wave
observations probe the classical dynamics of black holes with
unprecedented accuracy. The next frontier is to explore their quantum
properties, foremost among them Hawking radiation.

Hawking's prediction that black holes radiate thermally
\cite{Hawking:1974rv,Hawking:1975vcx}
established one of the deepest connections between gravity and quantum
mechanics, implying that sufficiently light black holes evaporate by
emitting all particle species whose masses lie below their
instantaneous Hawking temperature. Consequently, microscopic black
holes provide a natural laboratory for probing physics beyond the
Standard Model (SM), since any sufficiently weakly interacting particle
is expected to be produced through Hawking radiation once it becomes
kinematically accessible. Despite being one of the most profound
predictions of quantum field theory in curved spacetime, Hawking
radiation has yet to be observed experimentally.
Detecting its products would therefore provide a unique window into fundamental particle physics, black hole thermodynamics, and gravity at microscopic scales and extreme spacetime curvature.

High-energy neutrino astronomy has opened a new observational window onto the Universe. The IceCube Neutrino Observatory has firmly established the existence of an astrophysical neutrino flux extending from the TeV to the PeV scale \cite{IceCube:2013low,IceCube:2014stg}. More recently, KM3NeT \cite{KM3Net:2016zxf} reported the detection of the ultra-high-energy neutrino event KM3-230213A in the multi-PeV energy regime \cite{KM3NeT:2025npi}. These advances demonstrate that neutrino observatories are now probing previously unexplored energy regimes, where new physics may become accessible.
While an increasing fraction of this high-energy neutrino flux has been associated with extragalactic accelerators such as active galactic nuclei and blazars, its complete origin remains uncertain. At the same time, IceCube has reported evidence for a diffuse Galactic neutrino component associated with the Galactic plane \cite{IceCube:2023ame_Galactic}, motivating continued searches for new Galactic neutrino sources. In particular, mechanisms predicting characteristic energy spectra together with distinctive angular distributions constitute especially promising observational targets.

Neutron stars provide particularly favorable environments for exploring
the interplay between dark matter (DM), gravity, and particle physics (see,
e.g., Refs.~\cite{Dimopoulos:1982cz, Kolb:1982si} for pioneering works and
Ref.~\cite{Bramante:2023djs} for a recent review).
 Owing
to their enormous densities, deep gravitational potentials, and long
lifetimes, neutron stars efficiently capture dark matter particles from
the Galactic halo.
Over astrophysical timescales, the accumulated dark matter may become
self-gravitating and, depending on its microscopic properties, undergo
gravitational collapse. In asymmetric dark matter scenarios, where
annihilation is absent or highly suppressed, this collapse can lead to
the formation of microscopic black holes inside neutron stars. Their
subsequent evolution is governed by the competition between Hawking
evaporation and accretion from the surrounding medium. If the initial
black hole mass is sufficiently small, Hawking evaporation dominates and
the black hole evaporates before significant growth can occur. More massive black holes, on the other hand, may grow through
accretion and, under suitable conditions, ultimately destroy the host
neutron star. The observed existence of old neutron stars therefore
places  constraints on asymmetric dark matter models
\cite{Goldman:1989nd, Gould:1989gw, Kouvaris:2010jy, Kouvaris:2011fi,McDermott:2011jp, Kouvaris:2011gb, Bell:2013xk, Bramante:2013hn, Bramante:2013nma, Bramante:2017ulk, Garani:2018kkd, Dasgupta:2020dik, Dasgupta:2020mqg, Tinyakov:2021lnt, Garani:2021gvc, Bramante:2023djs, Bhattacharya:2023stq, Robles:2025dlv, Adarsha:2025jqs}.

The characteristic energy scale of Hawking radiation is determined by
the black hole temperature. For a non-rotating, electrically neutral
black hole,
\begin{equation}
\label{eq:TMBH}
kT_{\rm BH}
=
\frac{\hbar c^3}
{8\pi G M_{\rm BH}},
\end{equation}
or numerically,
$
kT_{\rm BH}
\simeq
1~{\rm PeV}
\left(
{10^{4}\ {\rm kg}}/
{M_{\rm BH}}
\right)$. 
Microscopic black holes with masses of order
$10^{4}$--$10^{6}\,$kg are therefore born with Hawking temperatures
naturally lying in the TeV--PeV range, making them attractive candidate
sources of ultra-high-energy neutrinos. As we show below, this  mass range is {\it not} assumed, but instead emerges naturally from the scenarios studied in this work. Neutrino telescopes therefore provide a particularly well-suited probe of their signatures.

\begin{figure}
    \centering
    \includegraphics[width=\linewidth]{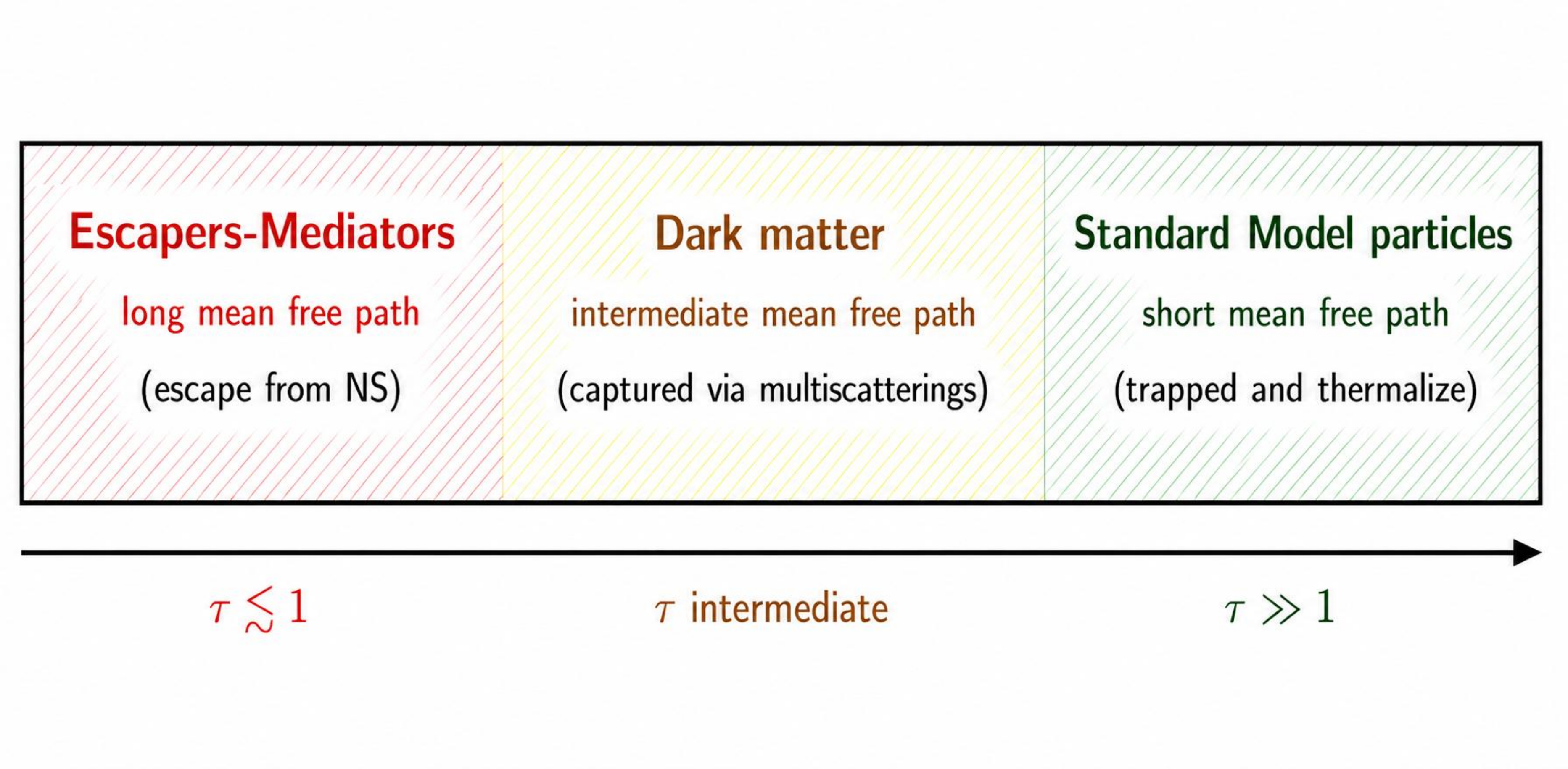}
    \caption{\small
Schematic illustration of the particle physics ingredients underlying
the present scenario. The three colored bands represent increasing
transparency of the neutron star, characterized by the optical depth $\tau$.
Standard Model particles are efficiently scattered and absorbed within
the stellar interior, while dark matter particles scatter and become
gravitationally captured. Long-lived, very feebly interacting particles
beyond the Standard Model escape the neutron star and subsequently decay
into high-energy neutrinos or other Standard Model particles outside the
stellar surface.
}
    \label{fig:Scenario}
\end{figure}

The evaporation of microscopic black holes naturally generates
extremely energetic particles. However, most Standard Model particles
are efficiently scattered and absorbed within the dense neutron star
interior before escaping. Their energy is instead deposited inside the
star, leading to a gradual heating of the neutron star surface, as
recently pointed out in Ref.~\cite{Saha:2025fgu}. Observable
high-energy messengers therefore require metastable, feebly
interacting particles that can traverse the stellar interior before
decaying into detectable Standard Model particles. Such states arise
naturally in many extensions of the Standard Model, including dark
photons, heavy neutral leptons, axion-like particles, scalar
mediators, and moduli fields. Throughout this work we adopt a
model-independent approach and consider a generic long-lived mediator
produced in Hawking radiation that escapes the neutron star and
predominantly decays into high-energy neutrinos. The underlying
mechanism is illustrated schematically in
Fig.~\ref{fig:Scenario}. Since the mediator properties enter primarily
through their lifetime and decay channels, the predicted neutrino
signal largely retains the characteristic features of the Hawking
spectrum.

In contrast to primordial black holes evaporating in vacuum, the
microscopic black holes considered here are generated continuously
through dark matter collapse inside neutron stars and are therefore
embedded in an extremely dense stellar medium. As a result, the
Standard Model Hawking products are absorbed before escaping,
naturally evading the stringent diffuse $\gamma$-ray and
charged particle constraints on evaporating primordial black holes,
first discussed in
Refs.~\cite{Page:1976wx,Carr:1976zz} and reviewed comprehensively in
Ref.~\cite{Carr:2020gox}. Related neutrino signatures from
dark matter induced microscopic black holes formed in the Sun and
Earth have also been investigated in Ref.~\cite{Acevedo:2020gro}.
Moreover, unlike primordial black holes, which spend most of their
lifetime as comparatively cold objects and emit TeV--PeV particles
only during their final stages of evaporation
\cite{Dave:2019epr,Anchordoqui:2025xug,Klipfel:2025jql}, the black
holes considered here are born hot. We find that their initial
Hawking temperatures are generally bounded from below at the TeV scale
and can readily extend into the PeV regime. Repeated dark matter
capture and subsequent gravitational collapse therefore lead to successive episodes of black
hole formation and evaporation, providing a
quasi-continuous source of high-energy neutrinos.

In this paper we investigate the complete sequence of dark matter
capture, gravitational collapse, microscopic black hole formation,
Hawking evaporation, and high-energy particle production 
inside neutron
stars. Particular emphasis is placed on the long-term evolution of the
captured dark matter cloud. We show that repeated black hole formation
and evaporation naturally lead to two qualitatively distinct regimes,
determined by the competition between the dark matter thermalization
time and the interval between successive collapse events. Besides the
fully thermalized regime previously discussed in the literature, we
identify a partially thermalized regime in which repeated evaporation
episodes continuously heat the dark matter cloud. The system evolves
toward a quasi-stationary configuration consisting of a hot dark matter
cloud embedded within the much cooler neutron star core. This
two-temperature structure governs the long-term evolution of the system
and gives rise to repeated microscopic black hole formation and
evaporation cycles.

Using this framework, we compute the time-integrated Hawking emission
and the associated secondary particle spectra, and derive the resulting
high-energy neutrino spectra from the decays of long-lived mediators.
We then calculate the expected neutrino fluxes focusing on the Galactic neutron star population, and the diffuse
extragalactic background. 
The predicted signal exhibits two distinctive
signatures: a broad, non-power-law neutrino spectrum whose peak is
determined primarily by the initial Hawking temperature of the
evaporating black hole, with a peak above
$\mathcal{O}(10),{\rm TeV}$, and a spatial distribution strongly
concentrated toward the Galactic Center, approximately tracing the
product of the neutron star distribution and the ambient dark matter
density. 
We compare the predicted signals with the diffuse
astrophysical neutrino flux measured by IceCube and estimate the
corresponding event rates in current and future neutrino telescopes.
Although the benchmark Galactic models considered here predict
subdominant event rates, significantly larger signals are obtained for
cuspy dark matter distributions. The proposed mechanism therefore
constitutes a well-defined multi-messenger target for future
observations, providing a direct connection between dark matter,
Hawking evaporation, compact objects, and high-energy neutrino
astronomy.

The paper is organized as follows.
In Sec.~II we discuss dark matter capture, thermalization, and
gravitational collapse inside neutron stars.
Section~III is devoted to repeated black hole formation, Hawking
evaporation, and the long-term evolution of the captured dark matter
cloud.
In Sec.~IV we derive the primary Hawking emission spectra and the
secondary neutrino spectra from mediator decays.
The resulting Galactic and extragalactic neutrino signals, together with
their observational implications, are presented in Sec.~V.
Finally, Sec.~VI contains our discussion and conclusions.

%

\section{Dark matter capture and accretion in neutron stars}

Dark matter particles approaching a compact star experience strong
gravitational focusing, which enhances the effective collision cross
section beyond the geometric value
\cite{Press:1985ug,Gould:1987ir,Goldman:1989nd}.
Consider a dark matter particle with asymptotic velocity $u_\chi$ and impact parameter $b$. 
The particle intersects the star if its periapsis lies within the stellar radius $R$. 
General relativistic effects enhance the focusing compared to the Newtonian treatment. In Schwarzschild spacetime, the critical impact parameter for a trajectory grazing the stellar surface is enhanced by the
relativistic factor 
$\tilde{\chi}\equiv \left(1-2GM/(Rc^2)\right)^{-1}$,
\begin{equation}
b_{\rm max}^2
=
\frac{R^2}{1-2GM/(Rc^2)}
\left(1+\frac{v_{\rm esc}^2}{u^2_\infty}\right)
\label{eq:bmax_GR}
\end{equation}
where $v_{\rm esc}^2 = {2GM}/{R}$ is the escape velocity at the stellar surface and $u_\infty$ is the asymptotic DM velocity far from the star. 
For neutron stars one has
$v_{\rm esc}\gg u_\infty$, corresponding to the strong focusing limit.
The effective collision cross section therefore becomes
$
\sigma_{\rm coll}
\simeq
\tilde\chi\pi R^2
{v_{\rm esc}^2}/{u_\infty^2},
$
with
$\tilde\chi\simeq1.5$--2 for typical neutron star compactness.  
The corresponding dark matter
mass flux through the neutron star is
\begin{equation}
\dot{M}_\chi
\simeq
\pi b_{\rm max}^2\,\rho_\chi\,u_\infty,
\end{equation}
where $\rho_\chi=n_\chi m_\chi$ is the ambient dark matter mass density.

\subsubsection{Capture condition}

In order for a dark matter particle crossing the neutron star to become
gravitationally bound, it must undergo at least one scattering and lose
sufficient energy such that its conserved energy, as measured by an
observer at infinity, falls below its rest-mass energy $m_\chi c^2$.
Assuming a uniform neutron density, the optical depth along a trajectory
with impact parameter $b$ is
\begin{equation}
\tau(b)
=
n_n\sigma_{\chi n}L(b),
\qquad
L(b)=2\sqrt{R_{\rm NS}^2-b^2},
\end{equation}
where $n_n$ is the neutron number density and $\sigma_{\chi n}$ is the
DM--neutron scattering cross section. Averaging over all isotropically
incident trajectories, the mean chord length through the star is
$\langle L\rangle=4R_{\rm NS}/3$, yielding
\begin{equation}
\tau
=
\frac{\sigma_{\chi n}}{\sigma_{\rm sat}},
\end{equation}
where
\begin{equation}
\sigma_{\rm sat}
\equiv
\frac{\pi R_{\rm NS}^2}{N_n}
\simeq
1.8\times10^{-45}\,{\rm cm}^2
\end{equation}
is the geometric saturation cross section and
$N_n\simeq M_{\rm NS}/m_n$ is the total number of neutrons in the star.
The optical depth $\tau$ therefore represents the mean number of
scatterings $N_{\rm sca}$ experienced by a DM particle during a single transit through
the stellar object.
For $\tau\ll1$ the neutron star is optically thin, whereas for
$\tau\gg1$ it is optically thick and essentially every particle
undergoes at least one scattering. 

For light ($m_\chi\sim m_n$) or
moderately heavy dark matter, one or a few scatterings are typically
sufficient for capture. In this regime, the condition
$\tau\gtrsim1$ is enough for the capture rate to approach the geometric
saturation limit.
For heavier DM particles, however, a single scattering event may not lead to sufficient energy loss \cite{Bramante:2017xlb}. 
In the limit of interest, $m_\chi \gg m_n$, the average value of the fractional energy loss per collision  reduces to
$
\epsilon_1 \simeq {2m_n}/{m_\chi}.
$
Assuming that the particle loses the same average fraction of energy in each collision, the cumulative energy loss after $N_{\rm sca}$ scatterings is
$\Delta E_{N_{\rm sca}}
=
E_K\left[1-(1-\epsilon_1)^{N_{\rm sca}}\right],
$
where $E_K=({\tilde{\chi}}^{1/2}-1)m_\chi c^2$ is the kinetic energy
of the incident DM particle at the stellar surface, as measured by a
local observer.
Requiring the energy lost during one transit $\Delta E_{N_{\rm sca}}$  to exceed the kinetic
energy at infinity we get the condition \cite{Goldman:1989nd}
\begin{equation}
2N_{\rm sca}\frac{m_n}{m_\chi}m_\chi c^2
\left[
1-\left(1-\frac{2GM}{Rc^2}\right)^{1/2}
\right]
\gtrsim
\frac12 m_\chi u_\infty^2 .
\label{eq:capture_condition}
\end{equation}
 in the limit $m_\chi \gg m_n$ and for $\epsilon_1 \ll 1$.

To ensure efficient capture of nearly all particles crossing the
neutron star, Eq.~\eqref{eq:capture_condition} is evaluated for a
velocity representative of the high-velocity tail of the
Maxwell--Boltzmann distribution. Specifically, we take
$u_\infty\simeq2\bar{u}_\infty$, where $\bar{u}_\infty$ is the mean halo
dark matter velocity. This choice also accounts for the uncertainty in
the dark matter velocity dispersion toward the Galactic Center
\cite{sofue2020rotationcurvemilkyway}. We then obtain
\begin{equation}
N_{\rm sca}
\gtrsim
N_{\rm req}
\simeq
3
\left(\frac{m_\chi}{10^6~{\rm GeV}}\right)
\left(\frac{\bar u_\infty}{220~{\rm km/s}}\right)^2.
\label{eq:Ncapture}
\end{equation}
Thus, for PeV-scale dark matter only a few scatterings are required for
capture, whereas for larger dark matter masses the required number of
scatterings grows linearly with $m_\chi$ and can become substantial.

\subsubsection{Heavy vs light dark matter}

Since $N_{\rm sca}\sim\tau$, efficient capture at the geometric saturation
limit for very heavy dark matter requires
\[
\tau \gtrsim N_{\rm req},
\]
which corresponds to a minimum scattering cross section
\begin{equation}
\sigma_{\rm req}
\simeq
N_{\rm req}\,
\sigma_{\rm sat},
\end{equation}
where $\sigma_{\rm req}$ denotes the minimum DM--neutron scattering
cross section required to reach the geometrically saturated capture rate
in the multi-scattering regime. When this condition is
satisfied, the vast majority of dark matter particles traversing the
neutron star lose sufficient energy to become gravitationally bound
during a single transit.
In this regime, the capture rate reaches the geometric saturation limit and is approximately given by
\begin{equation} \label{eq:Csat}
C_\chi^{\rm sat}
\simeq
\pi b_{\rm max}^2
\frac{\rho_\chi}{m_\chi}
u_\infty \,.
\end{equation}
For $\tau\ll  N_{\rm req}$ the
capture probability drops rapidly, as only particles in the
exponentially suppressed low-velocity tail lose sufficient energy to
become gravitationally bound during a single transit.

For a benchmark neutron star of mass $1.4\,M_\odot$, the accumulated
dark matter mass in the saturated capture regime is approximately
\begin{align}\label{eq:geometricaccretion}
M_{\rm acc}
\simeq
6\times10^{37}\,{\rm GeV}\;\,d_\chi\,\tilde{\sigma}_{\chi n}
\left(\frac{t}{\rm yr}\right),
\end{align}
where $d_\chi\equiv\rho_{\rm DM}/(10^3\,{\rm GeV\,cm^{-3}})$ parametrizes the
ambient dark matter density surrounding the neutron star. 
The
normalization corresponds to a benchmark dark matter density
$\rho_{\rm DM}=10^3\,{\rm GeV\,cm^{-3}}$, representative of dense
environments such as the Galactic Center or compact nuclear star
clusters.
For lower ambient densities, the accumulated mass scales
linearly with $\rho_{\rm DM}$. 
If $\chi$ constitutes only a fraction of the total dark matter,
Eq.~\eqref{eq:geometricaccretion} should be multiplied by
$\rho_\chi/\rho_{\rm DM}$.

\begin{figure}
    \centering
    \includegraphics[width=1.0\linewidth]{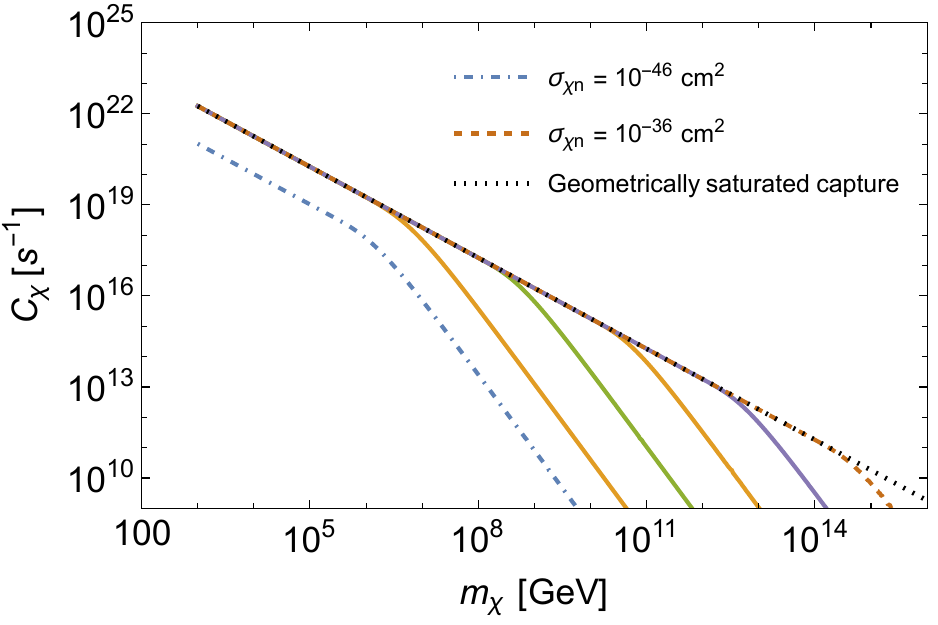}
    \caption{{\small Dark matter capture rate in a neutron star as a function of the DM mass
$m_\chi$ for cross sections $\sigma_{\chi n}/{\rm cm}^2=10^{-46}$, $10^{-44}$, $10^{-42}$, $10^{-40}$, $10^{-38}$, $10^{-36}$.
The calculation is performed in a multiscatter framework assuming
a uniform neutron star with $M_{\rm NS}=1.5\,M_\odot$ and $R_{\rm NS}=12\,{\rm km}$
 and averaging over a Maxwellian halo velocity distribution including gravitational focusing.
For $\sigma_{\chi n} \gtrsim \sigma_{\rm req}$ the capture rate approaches the
geometrically saturated value. At large $m_\chi$ the capture rate turns over because the fractional
energy loss per scattering decreases and an increasing number of
scatterings is required for capture.
}}
    \label{fig:CapR}
\end{figure}

For $N_{\rm req}=1$,
$\tilde\sigma_{\chi n}$ reduces to the usual capture-efficiency factor
\cite{Kouvaris:2010vv,McDermott:2011jp}.
For ultra-heavy dark matter ($N_{\rm req}>1$), we instead define
\begin{equation}
\tilde{\sigma}_{\chi n}
=
\frac{C_\chi(\tau)}
{C_\chi^{\rm sat}}, 
\end{equation}
where $C_\chi(\tau)$ is the capture rate obtained by averaging the
velocity dependent capture probability over the Maxwell--Boltzmann
velocity distribution, and $C_\chi^{\rm sat}$ is the
  the saturated capture rate, see Fig. \ref{fig:CapR}. 
For $\tau<N_{\rm req}$, the accumulated dark matter mass is expected to
be significantly suppressed relative to
Eq.~\eqref{eq:geometricaccretion}, as only the low-velocity tail of the
Maxwell--Boltzmann distribution can be efficiently captured. 

Of primary interest for the present work is the regime
$\tau\gtrsim N_{\rm req}$, in which
$\tilde{\sigma}_{\chi n}\simeq1$ and the capture rate approaches its
geometrically saturated value, for which
$
\dot{M}_{\rm acc}
\simeq
m_\chi C_\chi^{\rm sat}.
$
It is therefore important to examine whether the relatively large scattering cross
sections required for saturated capture remain compatible with current
direct detection constraints.

\subsubsection{Direct detection limits}

Current direct detection experiments also constrain the region of
parameter space in which geometrically saturated capture can be
realized. For ultra-heavy dark matter, $m_\chi\gg m_n$, the local dark
matter flux scales as $\rho_\chi/m_\chi$, while the nuclear recoil
spectrum becomes nearly independent of $m_\chi$. As a result, the
spin-independent direct detection bound weakens approximately linearly
with the dark matter mass,
\begin{equation} \label{eq:DD}
\sigma_{\chi n}^{\rm DD}
\simeq 10^{-36}\,{\rm cm}^2
\left(\frac{m_\chi}{10^{12}\,{\rm GeV}}\right),
\end{equation}
where the numerical coefficient is obtained by extrapolating the latest
LZ limits to the ultra-heavy mass regime \cite{LZ:2024psa, LZ:2024zvo}. Combining this result with the
condition for saturated capture,
\begin{equation}
\sigma_{\rm req}
\simeq
4\times10^{-39}\,{\rm cm}^2
\left(\frac{m_\chi}{10^{12}\,{\rm GeV}}\right)
\left(\frac{\bar u_\infty}{220~{\rm km/s}}\right)^2,
\end{equation}
one finds that the allowed parameter space is
\begin{equation}
\sigma_{\rm req}
\lesssim
\sigma_{\chi n}
\lesssim
\sigma_{\chi n}^{\rm DD}.
\end{equation}
Since both
$\sigma_{\rm req}$
and
$\sigma_{\chi n}^{\rm DD}$
scale approximately linearly with
$m_\chi$,
the allowed parameter space spans roughly
2.5 orders of magnitude over the entire mass range considered.

For a neutron star in the solar neighborhood, the corresponding dark
matter energy accretion rate is
$\dot E_\chi\sim10^{32\text{--}33}\,
{\rm GeV\,yr^{-1}}$,
corresponding to
$\dot M_\chi\sim10^6\,{\rm kg\,yr^{-1}}$.
In denser dark matter environments, such as the Galactic Center, this
rate can be orders of magnitude larger. Over sufficiently long
timescales, the accumulated asymmetric dark matter may become
self-gravitating and collapse into a microscopic black hole, providing
the starting point for the scenario explored in this work.

\section{Formation of black holes within neutron stars}

\subsection{Conditions for Black Hole Formation}

A self-gravitating dark matter core collapses to a black hole once its mass exceeds the corresponding Chandrasekhar-like stability limit. The critical mass depends on the particle statistics and on possible self-interactions.
For free (non-self-interacting) fermions and bosons they read  respectively,
\begin{equation} \label{eq:Chandra0}
M_{\rm Ch}^{(f)} \simeq 0.5 \frac{m_{\rm Pl}^3}{m_{f}^2}\,\,, \qquad \qquad M_{\rm Ch}^{(b)} \simeq 0.6 \,\frac{m_{\rm Pl}^2}{m_b}\,,
\end{equation}
where $m_{\rm Pl}=1.22\times10^{19}\,{\rm GeV}$ is the Planck mass and
$M_{\rm Pl}=m_{\rm Pl}/\sqrt{8\pi}$ is the reduced Planck mass.  We denoted the dark matter particle by $f$ in the fermionic case and by $b$ in the bosonic case. In the latter case the mass given by Eq. \eqref{eq:Chandra0} is called Kaup limit.  In what
follows, we collectively denote these collapse thresholds by
$M_{\rm Ch}$ whenever the distinction is not necessary.
 Setting $M_{\rm Ch}=10^{4} ~\mathrm{kg}\simeq 5.6 \times 10^{30}~\mathrm{GeV}$, which is a mass of interest,  yields
$m_f \approx
 1.8\times10^{13}~\mathrm{GeV}$ for free fermions and $m_b \approx  1.68\times10^{7}~\mathrm{GeV}$ for free bosons.

Repulsive self-interactions provide additional pressure support and
increase the collapse threshold. For fermions interacting through a
Yukawa mediator of mass $m_\varphi$ and coupling $g_\chi$, the pressure
receives an additional contribution, such that
$M_{\rm Ch}^{(f,{\rm self})}$ is increased by a factor
$\left(1+\alpha_{\rm rep}\right)$, where
$\alpha_{\rm rep}\approx g_\chi^2M_{\rm Pl}^2/m_\varphi^2$.
By contrast, attractive self-interactions reduce the effective
pressure support, thereby lowering the critical 
mass required for gravitational collapse. Consequently, microscopic
black holes may form for substantially lighter dark matter particles \cite{Kouvaris:2011gb}. In the present work we restrict our attention to non-interacting or
repulsively self-interacting dark matter, thereby maintaining a
largely model-independent framework.

In contrast to the free-boson case, the maximum stable mass for
interacting bosons scales as $M_{\rm Pl}^3/m_b^2$, similarly to the
fermionic Chandrasekhar limit, owing to the additional pressure
provided by repulsive self-interactions. For a quartic
self-interaction potential of the form
\begin{equation}
V(b)=\frac{\lambda}{2}|b|^4,
\end{equation}
the critical mass is given by the well-known
Colpi--Shapiro--Wasserman (CSW) limit,
\begin{equation} \label{Eq:ChandraInt}
M_{\rm Ch}^{(b,\mathrm{self})}
\simeq
0.06\,\sqrt{\lambda}\,
\frac{M_{\rm Pl}^3}{m_b^2}\,.
\end{equation}
Setting the benchmark black hole mass to
$M_{\rm BH}=M_{\rm Ch}^{(b,\mathrm{self})}\sim10^4~{\rm kg}$,
the corresponding benchmark bosonic dark matter mass is
\begin{equation}
m_b \approx
4.4\times10^{12}~{\rm GeV}\,\lambda^{1/4}.
\end{equation}

Observations of merging galaxy clusters constrain DM self-interactions through limits on the self-interaction cross section per unit mass. For the masses considered here, however, these bounds still allow non-perturbative values of $\lambda$. Whether the interacting expressions reduce to the non-interacting limit depends on the underlying microphysics. While symmetries can suppress self-interactions, extremely small couplings may be difficult to reconcile with sizable DM--nucleon interactions \cite{Bell:2013xk,Fan:2016rda}.

\subsubsection{Thermalization of captured dark matter}

The thermalization of captured DM proceeds through repeated scatterings
with the constituents of the neutron star core
\cite{Goldman:1989nd,
Bertone:2007ae, Kouvaris:2010vv, McDermott:2011jp,
Bertoni:2013bsa, Garani:2020wge, Bell:2023ysh}. In each collision the
DM particle transfers part of its kinetic energy to the medium,
gradually cooling until the typical energy transfer becomes comparable
to the neutron star temperature. Because the neutron star interior is
highly degenerate, only nucleons near the Fermi surface participate
efficiently in the scattering process, so Pauli blocking can
significantly increase the thermalization time.

The total thermalization time is  controlled by the final
low-energy scatterings, where the energy loss rate is smallest.
Consequently, the scaling of $t_{\rm th}$ with the DM mass changes
between the light and heavy DM regimes. Within the mass range
considered in Ref.~\cite{Bertoni:2013bsa} $t_{\rm th}$ decreases
with increasing DM mass.
This behavior is approximately described by the expression \cite{Bertoni:2013bsa, Garani:2020wge}
\begin{equation} \label{eq:tthermallight}
t_{\rm th} \simeq  10^4 {\rm yr} \frac{m_\chi m_{ n}}{(m_\chi+m_{ n})^2}\left(\frac{10^{-45} {\rm cm}^2}{\sigma_{\chi {n}}} \right) \left(\frac{10^5 {\rm K}}{T_{\rm NS}} \right)^2\,,
\end{equation}
which scales as $t_{\rm th}\propto m_\chi^{-1}$ for
$m_\chi\gg m_n$.
For very heavy dark matter, $m_\chi \gtrsim 10^8$ GeV, 
a different expression has been found \cite{Bell:2023ysh} 
\begin{equation}
t_{\rm th} \label{eq:tthermalheavy}
\simeq 10^{-3} {\rm yr}\,
\left(\frac{m_\chi}{10^9~{\rm GeV}}\right)
\left(\frac{2 \times 10^{-45}\,{\rm cm}^2}{\sigma_{\chi n}}\right)
\log\left(\frac{m_\chi}{T_{\rm NS}}\right)
\end{equation}

A compact DM core can form even while the DM population remains only
partially thermalized. Consequently, in the scenarios considered below,
the DM cloud may remain temporarily at a temperature different from that
of the surrounding neutron star medium. Since thermalization is governed
by DM--baryon scatterings, this qualitative picture applies equally to
fermionic and bosonic dark matter.

For our purposes, it is sufficient that the thermalization timescale satisfy
$
t_{\rm th} \lesssim {\rm Gyr},
$
a condition that holds throughout the ultra-heavy DM mass range considered
here. Since typically $
\sigma_{\chi n} \gg \sigma_{\rm sat},
$
DM undergoes many scatterings during each transit through the neutron
star, facilitating the formation of a compact central core.

Once thermalized, DM settles into the NS core.  A dilute, thermalized DM population has a Gaussian number density distribution $n_\chi(r)$ with a characteristic thermal radius
\begin{equation} \label{eq:rth}
r_{\rm th}
  = \left(\frac{3 k_{\rm B} T_{\rm NS}}{2\pi G\,\rho_{\rm NS}\, m_{\rm \chi}}\right)^{1/2}
\end{equation}
assuming a Maxwellian velocity distribution for the captured DM particles, and for $\rho_{\rm NS}= {3\times10^{17}~\mathrm{kg\,m^{-3}}}$.   
When the total DM mass enclosed within $r_{\rm th}$ exceeds that of hadronic matter in the same volume, $M_{\rm cloud}  > \frac{4\pi}{3}\rho_{\rm NS}r_{\rm th}^3$, the DM gravitational potential dominates locally and the system enters the self-gravitating regime. 
Applying the virial theorem then yields
\begin{align}
M_{\rm sg}=\sqrt{\frac{3 T^3_{\rm NS}}{\pi G^3 m^3_{\rm \chi}\rho_{\rm NS}}}\,.
\end{align}

A heavier DM particle  produces a more compact distribution and smaller  $M_{\rm sg}$.
A self-gravitating DM core of mass $M_{\rm sg}$ accumulates over a timescale
$
\Delta t_{\rm acc}^{\rm sg} \sim \times10^{11}~{\rm s}\,
\left({{\rm TeV}}/{m_{\chi}}\right)^{3/2}d_{\chi}^{-1}
$, when the capture rate reaches the geometric limit.
For all dark matter masses of interest in this work,
$\Delta t_{\rm acc}^{\rm sg}$ remains well below the typical neutron star
lifetime. 

The above discussion applies to both fermionic and bosonic DM, since thermalization is governed primarily by DM--baryon scattering. Bosonic DM may subsequently undergo Bose--Einstein condensation before gravitational collapse\footnote{
For free or sufficiently weakly self-interacting bosons, a
Bose--Einstein condensate (BEC) may form once the central density
becomes sufficiently large. The condensate occupies a radius much
smaller than the thermal radius $r_{\rm th}$, giving rise to a second,
more compact self-gravitating configuration. In this case, the relevant
criterion for BH formation is that the condensate mass exceeds the Kaup
limit. 
The discussion in
the main text therefore applies directly to fermions and to bosons with
appreciable repulsive self-interactions, while the free-boson BEC case
may follow a qualitatively different collapse sequence.}. Also, throughout this work we assume that dark matter--baryon
co-annihilations are negligible, so that the accumulated dark matter
population is not significantly depleted \cite{Bell:2013xk}.

\subsection{Evolution of Microscopic Black Holes}

Gravitational collapse begins once the captured DM cloud becomes
self-gravitating, $M_{\rm cloud}>M_{\rm sg}$. Whether this immediately
produces a black hole depends on the relation between the
self-gravitating mass and the Chandrasekhar-like collapse mass. If
$M_{\rm sg}<M_{\rm Ch}$ (or the CSW limit for self-interacting bosons),
additional dark matter must first accumulate before collapse occurs.
For the parameter space considered here,
\[
M_{\rm sg}<M_{\rm Ch/CSW},
\qquad
M_{\rm sg}>M_{\rm Kaup},
\]
so newly formed black holes are born with masses of order the
corresponding collapse mass.

The fate of a mini black hole formed inside a neutron star is determined by the competition between accretion, $\dot{M}_{\rm acc}\propto M_{\rm BH}^2$, and Hawking evaporation, $\dot{M}_{\rm evap}\propto -M_{\rm BH}^{-2}$. Hence, the rate of change of the newly born black hole mass 
can be written as
\begin{equation} \label{eq:bondHawk}
\frac{dM_{\rm BH}}{dt}
=
C_{\rm accr}\,M_{\rm BH}^2
-
\frac{c_{\rm evap}}{G^2 M_{\rm BH}^2}\, .
\end{equation}
For spherical accretion, the coefficient $C_{\rm accr}$ is known and is proportional to the nuclear matter density in the stellar core. Dark matter may also contribute to the accretion rate, depending on its local density and dynamics. The coefficient $c_{\rm evap}$ is dimensionless and depends on the number of particle species participating in Hawking radiation as well as on the black hole spin. 

There exists a critical black-hole mass above which accretion dominates
over Hawking evaporation. In the Bondi accretion regime this requires
\[
M_{\rm BH}^{\rm init}
<
m_{\rm Pl}
\left(
\frac{c_{\rm evap}}
{C_{\rm accr}}
\right)^{1/4},
\]
which corresponds to
\begin{equation}\label{eq:BHthresh}
M_{\rm BH}^{\rm init} < 5.7\times10^{36}\ {\rm GeV}
\simeq 10^{10}\ {\rm kg}
\end{equation}

The evolution of a microscopic black hole inside a neutron star may be
affected by the dense degenerate medium. However, Pauli blocking of
fermionic Hawking modes, neutron star rotation, and accretion-induced
heating introduce only modest corrections to the standard picture of
Bondi accretion and Hawking evaporation
\cite{Autzen:2014tza,Kouvaris:2013kra}. The dominant uncertainty instead
concerns the accretion mechanism itself: for black holes much smaller
than a nucleon, the Schwarzschild radius is well below microscopic
nuclear length scales, so the surrounding medium cannot be treated as a
continuous fluid and the applicability of the Bondi description becomes
questionable
\cite{Kouvaris:2013kra,Autzen:2014tza,Giffin:2021kgb}. As we show below,
once the core becomes gravitationally unstable, it rapidly collapses, adding its mass to the pre-existing central black hole,
providing an additional source of mass accretion.

Black holes heavier than the threshold in
Eq.~\eqref{eq:BHthresh} are expected to grow and ultimately destroy the
host neutron star \cite{Kouvaris:2011fi,McDermott:2011jp}, whereas
lighter ones evaporate (up to corrections from DM accretion). Inverting
this condition yields lower bounds on the DM particle mass for the
different collapse scenarios summarized in
Table~\ref{tab:DMbounds}.

\begin{figure}
    \centering
    \includegraphics[width=1.0\linewidth]{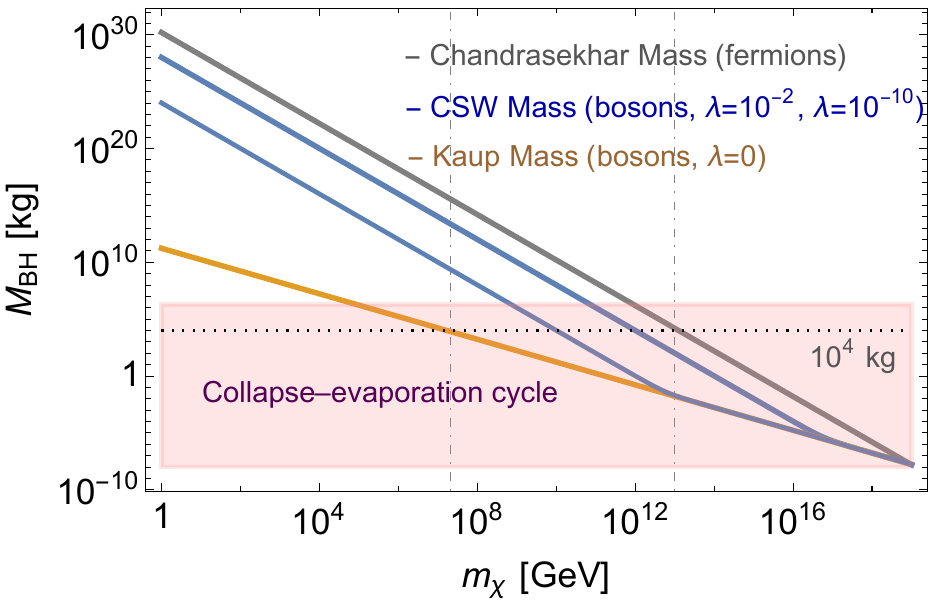}
    \caption{\small Initial black hole mass as a function of the dark matter particle mass
for fermionic and bosonic dark matter.
For bosonic dark matter, results are shown for three benchmark values of
the quartic self-interaction coupling:
$\lambda=0$, $\lambda=10^{-10}$, and $\lambda=10^{-2}$.
The shaded pink region satisfies the evaporation condition,
Eq.~\eqref{eq:MevapDM}, assuming geometrically saturated dark matter
capture, and therefore corresponds to repeated black hole evaporation
events.
The horizontal dashed line marks the benchmark black hole mass
$M_{\rm BH}=10^{4}\,{\rm kg}$. 
}
    \label{fig:BHrange}
\end{figure}

 The time required to accumulate a dark matter core of mass
$M_{\rm BH}$ inside the neutron star is $
\Delta t_{\rm acc}
=
{M_{\rm BH}}/{m_\chi C_\chi}= M_{\rm BH}/\dot{M}_{\rm acc}$.
Once this core becomes gravitationally unstable, it rapidly collapses,
and its mass is incorporated into the pre-existing central mini black hole. 
Numerically,
\begin{align}
\Delta t_{\rm acc}(M_{\rm BH})
\simeq
3\,{\rm s}
\left(\frac{M_{\rm BH}}{10^4\,{\rm kg}}\right)
\left(\frac{\dot M_{\rm acc}}
{6\times10^{37}\,{\rm GeV\,yr}^{-1}}\right)^{-1}.
\end{align}
 Taking $M_{\rm BH}=M_{\rm Ch}$, repeated evaporation events require that the existing black hole evaporate before the subsequent Chandrasekhar-mass dark matter core collapses and merges with it. This implies the condition
\begin{equation} \label{eq:evap_cond}
t_{\rm evap}(M_{\rm Ch}) < \Delta t_{\rm acc}^{\rm Ch}.
\end{equation}
Equation~\eqref{eq:evap_cond} provides a sufficient criterion for repeated
black hole evaporation events. If Hawking heating delays the formation of
the next collapsing core, the interval between successive collapse events
may exceed $\Delta t_{\rm acc}^{\rm Ch}$. In this case, the system is
expected to relax toward a quasi-stationary state in which Hawking heating
maintains a hot, partially thermalized, self-gravitating dark matter cloud,
while repeated gravitational collapse continues, as discussed in the
following section. Since $\Delta t_{\rm acc}^{\rm Ch}$ depends on both the
DM mass and the ambient DM density, Eq.~\eqref{eq:evap_cond} translates
into a lower bound on the DM particle mass that can be more restrictive
than that obtained from baryonic accretion alone.
Using
\begin{equation} \label{eq:tevap}
t_{\rm evap}(M_{\rm BH})
\simeq
8.4\times10^{-5}\,{\rm s}
\left(\frac{M_{\rm BH}}{10^4\,{\rm kg}}\right)^3,
\end{equation}
Eq.~\eqref{eq:evap_cond} implies that the black hole evaporates before
the next Chandrasekhar-mass dark matter core forms, provided that
\begin{equation}
\label{eq:MevapDM}
M_{\rm BH}
\lesssim
1.9\times10^6\,{\rm kg}\,
\left(
\frac{\dot M_{\rm acc}}
{6\times10^{37}\,{\rm GeV\,yr}^{-1}}
\right)^{-1/2},
\end{equation}
for which the Hawking evaporation and dark matter accumulation
timescales become comparable. The corresponding initial Hawking
temperature satisfies
\begin{equation}
kT_{\rm BH}\gtrsim5.3~{\rm TeV},
\end{equation}
which translates into the lower bounds
\begin{align}
m_\chi^{\rm Kaup} &\gtrsim 9\times10^4~{\rm GeV}, \nonumber\\
m_\chi^{\rm CSW} &\gtrsim 3\times10^{11}\lambda^{1/4}~{\rm GeV}, \nonumber\\
m_\chi^{\rm Ch} &\gtrsim 9\times10^{11}~{\rm GeV}.
\end{align}
assuming the geometric saturation capture rate,
$\dot M_{\rm acc}=\dot M_{\rm acc}^{\rm sat}$.
The corresponding limits for the different collapse scenarios are
summarized in Table~\ref{tab:DMbounds}. Whenever
Eq.~\eqref{eq:MevapDM} is satisfied, the neutron star is expected to
undergo repeated cycles of dark matter collapse, microscopic black hole
formation, and complete Hawking evaporation.

\begin{table}[t]
\centering
\begin{tabular}{lcc}
\hline
\textbf{ADM Model} &
\multicolumn{2}{c}{\textbf{Lower bound on $m_\chi$ (GeV)}} \\
\cline{2-3}
&
\small NS accretion &
\small DM replenishment \\
\hline

\small Fermionic (Chandrasekhar)
&
$1.3\times10^{10}$
&
$9.2\times10^{11}$
\\[4pt]

\small Bosonic (CSW, $\lambda=1$)
&
$4.5\times10^{9}$
&
$3.2\times10^{11}$
\\[4pt]

\small Bosonic (CSW, $\lambda=10^{-10}$)
&
$1.4\times10^{7}$
&
$1.0\times10^{9}$
\\[4pt]

\small Bosonic (CSW, $\lambda=10^{-20}$)
&
$4.5\times10^{4}$
&
$3.2\times10^{6}$
\\
\hline
\end{tabular} 
\caption{\small
Lower bounds on the dark matter particle mass, $m_\chi$, in GeV for
representative asymmetric dark matter models. The ``NS accretion''
column gives the minimum mass required for Hawking evaporation to
overcome neutron star matter accretion (Eq.~\eqref{eq:BHthresh}). The
``DM replenishment'' column gives the lower bound obtained by requiring
the black hole to evaporate before a new Chandrasekhar-mass dark matter
core is accumulated and collapses (Eq.~\eqref{eq:MevapDM}). A benchmark
saturated dark matter accretion rate of
$\dot{M}_{\rm acc}=6\times10^{37}\,{\rm GeV\,yr^{-1}}$
is assumed throughout.
} \label{tab:DMbounds}
\end{table}

\section{Black Hole Evaporation Cycles and Dark Matter Cloud Rethermalization} \label{sec:Per}

\subsection{Dark Matter Cloud Heating and Rethermalization}

Each cycle of microscopic black hole formation and evaporation injects
energy into the neutron star through two channels: Hawking radiation
and the gravitational binding energy released during the collapse of the
dark matter core. While the heating of neutron star matter is
negligible, Hawking radiation can efficiently transfer energy to the
surrounding dark matter cloud. If dark sector particles are emitted and
subsequently decay or scatter into the ambient dark matter, the cloud
may be partially or completely dispersed.

The evaporation of a microscopic black hole deposits energy into the
surrounding neutron star medium. Assuming equipartition, the
corresponding increase in the baryonic temperature is
\begin{equation}
\Delta T_{\rm n}
\sim
\frac{M_{\rm BH}}{N_{\rm n}}.
\end{equation}
For Chandrasekhar-scale collapse masses, one finds
$
\Delta T_{\rm n}\ll T_{\rm NS}$,
and the associated heating is therefore entirely negligible compared
with the typical core temperature of a neutron star. Moreover, the heat
released near the stellar center is redistributed on a timescale shorter
than, or at most comparable to,
$\Delta t_{\rm acc}^{\rm Ch}$.
This conclusion is further supported by the enormous heat capacity of
neutron star matter and its rapid thermal diffusion
\cite{Lattimer:1994glx,Shapiro:1983du},
even in the presence of nucleon superfluidity. Thus, the evaporation of microscopic black holes does not appreciably modify the global thermal state of the neutron star.

Although each individual event produces negligible heating, repeated
evaporation cycles may maintain a minimum neutron star surface
temperature, analogously to dark matter annihilation or kinetic heating
\cite{Kouvaris:2007ay,Baryakhtar:2017dbj}.

Upon BH evaporation, a fraction of the Hawking radiation may be
transferred to the dark sector coupled to the DM  and subsequently deposited into a local
population $N_\chi^{}$ of ambient DM particles surrounding the BH. If
a fraction $f_{d|H}$ of the Hawking luminosity is emitted into dark sector
degrees of freedom, the total energy injected into the ambient DM cloud is
approximately
$
E_{\rm dep}
\sim
f_{d|H}\,M_{\rm BH}.
$
The corresponding temperature increase can be estimated as

\begin{equation} \label{eq:Tdisrupt}
\Delta T_{\chi}
\sim
\frac{f_{d|H}\,M_{\rm BH}}
     {N_\chi^{}}.
\end{equation}

Unlike the baryonic component, whose enormous heat capacity suppresses any
appreciable temperature increase, the ambient DM cloud contains a comparatively
small number of particles. Consequently, even a modest coupling of Hawking
radiation to the dark sector can result in a very large energy injection per DM
particle. In particular, the deposited energy can greatly exceed the thermal
energy stored in the cloud 
$
N_\chi^{}\,T_{\rm NS}\,
$
leading to the evaporation and dispersal of the thermalized DM configuration.

Since the energy is deposited on the evaporation timescale,
\[
t_{\rm evap}\ll t_{\rm th},
\]
where $t_{\rm th}$ is the characteristic DM thermalization time
(Eq.~\eqref{eq:tthermalheavy}), the ambient DM cloud is driven far from
thermal equilibrium before DM--baryon or DM self-scattering processes
can redistribute the injected energy. Whenever
\[
f_{d|H}\,M_{\rm BH}
\gg
N_\chi\,T_{\rm NS},
\]
the deposited Hawking energy exceeds the thermal binding energy of the
ambient DM cloud, causing it to disperse throughout the neutron star
interior. Consequently, unless $f_{d|H}=0$, each evaporation episode
destroys the pre-existing thermalized (or partially thermalized) DM
configuration. The heated DM particles must subsequently rethermalize
and reaccumulate before another compact core can form. Since dark matter
continues to be captured throughout this process, its total abundance
inside the neutron star steadily increases, while the central DM
distribution may instead consist of a hot cloud with
$T_\chi\gg T_{\rm NS}$ rather than a thermalized sphere.

Let us compare two key timescales governing the evolution of the dark
matter core: the thermalization time $t_{\rm th}$ and the Chandrasekhar
mass accumulation time $\Delta t_{\rm acc}^{\rm Ch}$. When
\begin{equation} \label{eq:thermcond}
    t_{\rm th}<\Delta t_{\rm acc}^{\rm Ch}
\end{equation}
the captured DM fully thermalizes before a Chandrasekhar mass is
accumulated, and gravitational collapse proceeds from a thermalized
cloud.

Using $t_{\rm th}$ from Eq. \eqref{eq:tthermalheavy} for heavy dark matter $m_{\chi}\gtrsim 10^8$ GeV (otherwise Eq. \eqref{eq:tthermallight})
the condition \eqref{eq:thermcond}$ $ can be written as
\begin{equation}
\sigma_{\chi n}
\gtrsim
\sigma_{\chi n}^{\rm th},
\end{equation}
where $\sigma_{\chi n}^{\rm th}$ is the (fast) thermalization cross section defined by
$t_{\rm th}=\Delta t_{\rm acc}^{\rm Ch}$. For a typical neutron star core temperature
$T_{\rm NS}\sim10^{5}$--$10^{6}\,$K, the logarithmic factor varies only
weakly with the dark matter mass and is approximately
$\log(m_\chi/T_{\rm NS})\simeq45$ for
$m_\chi\sim10^{12}\,$GeV. For fermionic dark matter the thermalization cross section therefore becomes
\begin{equation}
\sigma_{\chi n,f}^{\rm th}
\simeq
6\times10^{-39}\,{\rm cm}^2
\left(\frac{m_f}{10^{12}\,{\rm GeV}}\right)^3
\left(
\frac{\dot M_{\rm acc}}
{\dot M_{\rm acc}^{\rm sat}}
\right).
\end{equation}
where $\dot M_{\rm acc}^{\rm sat}=m_\chi C_\chi^{\rm NS,sat}$ is the
dark matter accretion rate in the saturation regime.
For $m_f\sim10^{12}\,{\rm GeV}$, $\sigma_{\chi n,f}^{\rm th}$ is comparable to the geometric saturation cross section, whereas it increases rapidly with the DM mass.
For bosonic dark matter with repulsive quartic self-interactions,
\begin{equation}
\sigma_{\chi n,b}^{\rm th}
\simeq
5\times10^{-38}\,{\rm cm}^2\,
\lambda^{-1/2}
\left(\frac{m_b}{10^{12}\,{\rm GeV}}\right)^3
\left(
\frac{\dot M_{\rm acc}}
{\dot M_{\rm acc}^{\rm sat}}
\right).
\end{equation}

\begin{figure}
    \centering
    \includegraphics[width=1.0\linewidth]{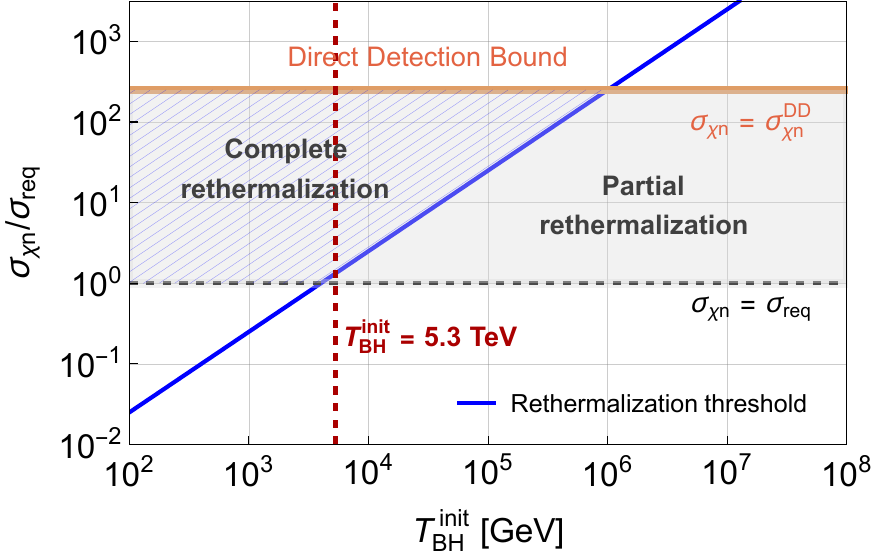}
    \caption{{\small 
Parameter space for repeated microscopic black hole formation in terms
of the initial Hawking temperature and the DM--nucleon scattering cross
section, normalized to the minimum cross section required for
geometrically saturated capture,
$\sigma_{\rm req}$.
The blue curve separates the fully and partially thermalized regimes,
assuming Eq.~\eqref{eq:tthermalheavy} for the thermalization time.
The shaded region satisfies both the geometrically saturated capture
condition and current direct detection constraints, while the hatched
region corresponds to complete rethermalization of the dark matter cloud
before the next collapse. The vertical dashed line indicates the minimum
initial Hawking temperature,
$T_{\rm BH}^{\rm init}\simeq5.3~{\rm TeV}$,
for which the black hole evaporates before another Chandrasekhar-mass
dark matter core forms. 
}}
    \label{fig:rethermal}
\end{figure}

Requiring the thermalization cross section to remain below the current
direct detection upper bound,
\begin{equation}
\sigma_{\chi n,f}^{\rm th}
<
\sigma_{\chi n}^{\rm DD},
\end{equation}
places upper limits on the fermionic and bosonic dark matter masses.
These upper limits translate into lower bounds on the initial black
hole mass. Since
$M_{\rm Ch}^{(f)}\propto m_f^{-2}$ for fermionic dark matter and
$M_{\rm Ch}^{(b,{\rm CSW})}\propto\sqrt{\lambda}/m_b^2$ for bosonic
dark matter with repulsive quartic self-interactions, one finds
\begin{equation}
M_{\rm BH}
\gtrsim
10^4\,{\rm kg}
\left(
\frac{\dot M_{\rm acc}}
{\dot M_{\rm acc}^{\rm sat}}
\right),
\end{equation}
where we have used
$\log(m_\chi/T_{\rm NS})\simeq45$ for
$m_\chi\sim10^{12}\,{\rm GeV}$ and
$T_{\rm NS}\sim10^5$--$10^6\,{\rm K}$.
The corresponding initial Hawking temperature satisfies
\begin{equation}
kT_{\rm BH}
\simeq
1~{\rm PeV}
\left(
\frac{10^4\,{\rm kg}}{M_{\rm BH}}
\right)
\lesssim
1~{\rm PeV}
\left(
\frac{\dot M_{\rm acc}}
{\dot M_{\rm acc}^{\rm sat}}
\right)^{-1}.
\end{equation}
 This bound, as follows from Eq.~\eqref{eq:tthermalheavy},  is necessary,
but not sufficient, for realizing the fully thermalized collapse
regime. In particular, assuming geometric saturation of the capture
rate,
$\dot M_{\rm acc}=\dot M_{\rm acc}^{\rm sat}$,
the combined requirements
$t_{\rm th}<\Delta t_{\rm acc}^{\rm Ch}$ and
$\sigma_{\chi n}^{\rm th}<\sigma_{\chi n}^{\rm DD}$ imply
\begin{equation}
M_{\rm BH}\gtrsim10^4\,{\rm kg},
\qquad
kT_{\rm BH}\lesssim1~{\rm PeV}.
\end{equation}
The thermalization cross section therefore delineates two qualitatively
distinct collapse regimes. The precise location of this transition is,
however, only approximate, since the available estimates of
$t_{\rm th}$ rely on different approximations and are valid in
different regions of parameter space. For
$\sigma_{\chi n}>\sigma_{\chi n}^{\rm th}$,
thermalization is faster than Chandrasekhar-mass accumulation, and
collapse proceeds from a fully thermalized cloud. Conversely, for
$\sigma_{\chi n}<\sigma_{\chi n}^{\rm th}$,
gravitational collapse occurs before complete thermalization and
therefore originates from a partially thermalized cloud. These two
regimes are illustrated in Fig.~\ref{fig:rethermal}. In the remainder
of this section, we focus on the partially thermalized regime.

\subsection{Formation of a partially thermalized  dark matter cloud}

\subsubsection{Black hole evaporation and partial rethermalization}
For
$
\sigma_{\chi n}
\lesssim
\min\!\left[
\sigma_{\chi n}^{\rm th}(m_\chi),
\sigma_{\chi n}^{\rm DD}(m_\chi)
\right]
$
(and
$\sigma_{\rm req}(m_\chi)\lesssim\sigma_{\chi n}$,
the regime of interest in this work), the dark matter cloud reaches
the Chandrasekhar mass before complete thermalization. In this regime,
the energy released by each black hole evaporation event heats and
partially disperses the thermalized dark matter core, while the
surrounding cloud rethermalizes on a timescale longer than the mass
accumulation time. Consequently, the
subsequent formation of a new self-gravitating core, and hence the next
black hole formation event, is no longer immediate.

A new dark matter core can collapse only after the captured dark matter cloud
has cooled and recontracted. The core becomes stable against further
thermal disruption once its thermal energy is comparable to the energy
deposited by Hawking evaporation, defining a characteristic equilibrium
temperature of the dark matter core.
Equating the temperature increase \eqref{eq:Tdisrupt}
to its intrinsic temperature $T_{\chi}$ gives the condition
\begin{equation}
m_{\chi}\frac{f_{d|H} M_{\rm Ch}}{M_{\chi}} \sim T_{\chi}.
\end{equation}
Hence,  the DM particles are in thermal equilibrium 
at temperature $T_{\chi}$, which is different than the internal temperature of the star and form a virialized sphere of radius $r_{\chi}$.
Because $T_{\chi} > T_{\rm NS}$, the thermal radius of the DM cloud expands to
\begin{align}
r_\chi = r_{\rm th} \sqrt{\frac{T_{\chi}}{T_{\rm NS}}}.
\end{align}
The maximum, and therefore equilibrium, cloud mass is the amount required for
self-gravitation,  
\begin{equation} \label{Eq:Mhalo}
M_{\rm sg,\chi}
=\frac{4\pi}{3}\rho_{\rm NS}r_{\rm th}^3
\left(
\frac{T_\chi}{T_{\rm NS}}
\right)^{3/2}.
\end{equation}
After the first evaporation event the disrupted cloud cools while
additional dark matter continues to accumulate. Repeated
collapse--evaporation cycles therefore drive the system toward a
quasi-stationary cloud characterized by an equilibrium temperature
$T_\chi$ and a self-gravitating mass $M_{\rm sg,\chi}$. In this state,
the energy released by each evaporation event constitutes only a small
perturbation to the cloud's total thermal energy, temporarily halting
its contraction without dispersing the cloud.

The corresponding equilibrium temperature for fermionic and bosonic
dark matter is
\begin{equation}
T_{\chi}\sim c_{f/b}\,10^{11}~{\rm K}\,
f_{d|H}^{2/5}
\left(\frac{m_{\chi}}{\rm GeV}\right)^{1/5}
\left(\frac{\rho_{\rm NS}}{\rho_{\rm NS}^0}\right)^{-2/5},
\end{equation}
where $c_f\simeq20$ for fermions,
$c_b\simeq6.5\,\lambda^{1/5}$ for self-interacting bosons, and
$\rho_{\rm NS}^0=3\times10^{17}~{\rm kg\,m^{-3}}$.
As shown in Fig.~\ref{fig:TNScore}, this temperature exceeds the
initial thermalized DM temperature by several orders of magnitude.

The corresponding self-gravitating cloud mass is
\begin{equation}
M_{{\rm sg},\chi}
\sim
4\times10^{15}\,{\rm kg}\;
C_{f/b}\,
f_{d|H}^{3/5}
\left(\frac{10^{12}\,{\rm GeV}}{m_\chi}\right)^{6/5}
\left(\frac{\rho_{\rm NS}}{\rho_{\rm NS}^0}\right)^{-11/10}
\end{equation}
where $C_f\simeq8.5$ for fermions and
$C_b\simeq1.5\,\lambda^{3/10}$ for self-interacting bosons.
Consequently, complete thermalization is not required for
 gravitational collapse: the dark matter cloud can become self-gravitating
even when its radius substantially exceeds the thermal radius
$r_{\rm th}$.
For illustration, a fermionic dark matter particle with
$m_\chi=10^{13}\,{\rm GeV}$ has
$r_{\rm th}\simeq4~\mu{\rm m}$,
whereas the corresponding hot cloud reaches
$r_\chi\simeq4.5~{\rm cm}$, illustrating that repeated black hole
evaporation inflates the dark matter distribution far beyond thermal
equilibrium. Assuming geometrically saturated capture, such a
self-gravitating cloud forms in approximately
${\cal O} (10^{4})\,$yr.
This formation time also provides an order of magnitude estimate of the
duration of the transient evolution toward the quasi-stationary regime.
Since it is much shorter than the typical age of an old neutron star,
such systems are expected to spend most of their lifetime in the
quasi-stationary state described above.

\subsubsection{Gravitational compression and BH formation}

After a time interval $\Delta t_{\rm acc}^{\rm Ch}$, an amount of dark
matter equal to the Chandrasekhar mass $M_{\rm Ch}$ has been added to
the partially thermalized DM cloud, while an equal mass has accumulated
in the central region and collapsed into a black hole. 

As DM capture continues, the accumulated cloud eventually becomes
self-gravitating. Owing to its centrally peaked density profile, the
innermost region becomes unstable first and undergoes gravothermal
contraction, while the outer cloud remains approximately thermal.
During this evolution, the gravitational binding energy released by the
contracting core is transported to the surrounding DM cloud and neutron
star medium, regulating the contraction in close analogy with classical
gravothermal collapse
\cite{Antonov1962,LyndenBellWood1968,Chavanis2002,BinneyTremaine2008}.

For fermionic DM, contraction leads to the formation of a degenerate
core that eventually reaches the Chandrasekhar mass. For bosonic DM,
collapse proceeds until the Kaup or CSW stability limit is exceeded.
In the special case of non interacting bosons, a Bose--Einstein
condensate may form before collapse, providing an even more compact
configuration that rapidly collapses once its critical mass is
exceeded.

The approach to these stability limits is controlled by the preceding
gravothermal evolution. The gravothermal contraction
proceeds on a timescale comparable to the thermalization time,
$t_{\rm gt}\sim t_{\rm th}$,
whereas the final gravitational collapse occurs on the much shorter
free-fall timescale,
$t_{\rm ff}\lesssim{\cal O}(1)\,{\rm s}$.
Thus, the bottleneck for microscopic black hole formation is the
gravothermal evolution rather than the collapse itself.

\subsubsection{BH Evaporation Cycles and Two-Temperature Structure}

In the fully thermalized regime
($t_{\rm th}<\Delta t_{\rm acc}^{\rm Ch}$), the interval between
successive BH formation events is simply the time required to
accumulate a Chandrasekhar mass of DM,
$
\Delta t_{\rm cyc}\simeq
\Delta t_{\rm acc}^{\rm Ch}$. 
The partially thermalized regime
($t_{\rm th}>\Delta t_{\rm acc}^{\rm Ch}$) exhibits a qualitatively
different transient evolution. Following each BH evaporation event,
the central DM distribution is heated and partially disrupted.
Nevertheless, DM capture and thermalization continue uninterrupted,
leading to the formation of an increasingly massive, partially
thermalized cloud. As the cloud mass grows, an ever larger fraction of particles occupies
low-energy orbits, forming a reservoir that continuously feeds the
formation of successive collapsing cores.

Once the cloud reaches the quasi-stationary configuration discussed in
the previous subsection, the energy deposited by each BH evaporation
event becomes only a small perturbation to the total thermal energy of
the cloud. Consequently, the relaxation time $t_{\rm th}$ no longer
determines the interval between successive collapses. Instead,
thermalization continuously maintains the low-energy reservoir, while
each new collapse is triggered by the accumulation of an additional
mass $M_{\rm Ch}$ in the central region. The recurrence time therefore asymptotically approaches
\begin{equation}
\Delta t_{\rm cyc}
\simeq
\Delta t_{\rm acc}^{\rm Ch}
=
\frac{M_{\rm Ch}}{\dot M_{\rm acc}},
\label{eq:dtcyc}
\end{equation}
with corrections arising from the finite post-evaporation gravothermal
relaxation of the cloud. These corrections are expected to become
progressively smaller as the quasi-stationary cloud grows and the energy deposited by each BH evaporation event becomes only a
small perturbation to the total thermal energy of the cloud.

Thus, the thermalization time governs only the initial transient
establishment of the quasi-stationary cloud, whereas the long-term
periodicity of BH formation and evaporation is controlled by the DM
accumulation rate. Since
$\Delta t_{\rm acc}^{\rm Ch}\propto(d_\chi m_\chi)^{-1}$,
BH formation cycles become less frequent in DM-poor environments,
such as the solar neighborhood, and more frequent in DM-rich regions,
such as globular clusters or the Galactic Centre.

The partially thermalized regime may also admit a runaway scenario in
which the rebuilding of the central core outpaces Hawking evaporation,
leading to progressively more massive black holes and eventual neutron
star destruction. However, for the benchmark black hole masses considered here,
$t_{\rm evap}\lesssim0.1\,{\rm s}$, given by
Eq.~\eqref{eq:tevap}, remains many orders of magnitude shorter than
the corresponding accretion, thermalization, and gravothermal
timescales. Our analysis therefore favors a quasi-stationary cloud that
repeatedly feeds the formation of microscopic black holes rather than
runaway growth.

\begin{figure}
    \centering
    \includegraphics[width=1.0\linewidth]{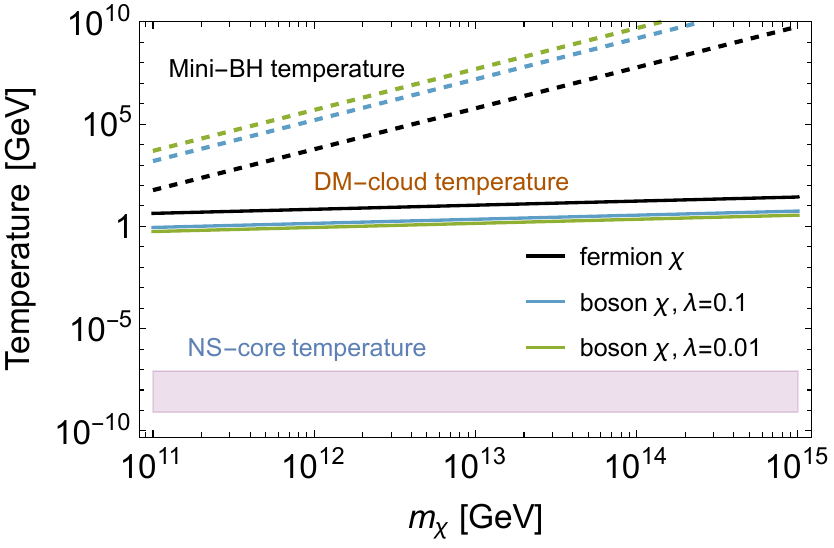}
    \caption{{\small Equilibrium temperature of the quasi stationary
partially thermalized DM cloud (solid curves), corresponding to the regime $t_{\rm th}>\Delta t_{\rm acc}^{\rm Ch}$, together with the initial Hawking temperature of the promptly evaporating black hole (dashed curves), shown as functions of the fermionic or bosonic dark matter particle mass. The horizontal lines denote representative neutron star core temperatures. The hierarchy $T_{\rm BH}\gg T_{\rm cloud}\gg T_{\rm NS}$ illustrates that Hawking evaporation heats the DM cloud much more efficiently than the surrounding neutron star medium. }}
    \label{fig:TNScore}
\end{figure}

In the core of a neutron star hosting a DM cloud and a central evaporating
BH, the thermal state is intrinsically out of equilibrium. Because black hole evaporation events recur on the timescale
$\Delta t_{\rm cyc}$, they act as an approximately continuous heat
source.  However, DM and
baryons exchange energy only through the weak DM--baryon interactions,
so thermal equilibration between the two sectors occurs on the 
 timescale $t_{\rm th}$ which can be longer than $\Delta t_{\rm cyc}$. Consequently, energy deposited into the DM component remains largely confined within that sector, allowing it to maintain a temperature higher than that of the surrounding baryonic medium.  By contrast, baryons rapidly
redistribute the injected heat through strong nuclear interactions,
maintaining an approximately uniform temperature. The neutron star core therefore naturally develops a two-temperature structure whenever $t_{\rm th}>\Delta t_{\rm cyc}$, consisting of
a hot DM cloud embedded in a cooler, strongly thermalized baryonic
background, as illustrated in Fig.~\ref{fig:TNScore}.

\subsection{Observational Implications}

The establishment of this periodic cycle marks the point at which the
DM-induced BH dynamics become predictable and directly linked to
observable consequences. Each collapse--evaporation episode injects
energy into the surrounding medium. Consequently, DM-induced heating of
neutron stars, through both kinetic heating and energy deposition from
BH evaporation, may contribute to the surface temperature, particularly
for old neutron stars with low intrinsic temperatures, as recently
discussed in Ref.~\cite{Saha:2025fgu}.

Each evaporation event produces a well defined Hawking spectrum of
Standard Model and dark sector particles. For an individual neutron
star the recurrence time is set by
$\Delta t_{\rm cyc}$,
while the burst energetics are determined by
$M_{\rm Ch}$. For a population of neutron stars, the superposition of
many independent events effectively forms a continuous source.

In what follows, we examine how these repeated evaporation events give rise to potentially observable signals in neutrino telescopes. Conversely, their non-observation can be used to constrain the dark matter mass, self-interactions, and ambient dark matter density.

\section{Particle emission from Black Hole  evaporation}

\subsection{Primary Spectrum}

The microscopic black holes considered here have Schwarzschild radii
\begin{equation}
r_s=\frac{2GM_{\rm BH}}{c^2}\simeq
1.5\times10^{-23}\,{\rm m}
\left(\frac{M_{\rm BH}}{10^4\,{\rm kg}}\right),
\end{equation}
which are $8$--$10$ orders of magnitude smaller than the nucleon size ($r_{\rm n}\sim10^{-15}\,$m), and many orders of magnitude smaller than the characteristic scales over which the neutron star medium varies. Consequently, the horizon probes only an infinitesimal region of spacetime, where the geometry is well approximated by the local Schwarzschild solution. 
Hawking radiation originates within a region of size
$\mathcal{O}(r_s)$ around the horizon and is therefore determined
primarily by the local spacetime geometry. Although the dense neutron
star medium can modify the propagation of the emitted particles, for
example through Pauli blocking, these effects produce only modest
corrections to the evaporation rate. The standard Hawking spectrum is
therefore expected to remain a good approximation.

Let us assume that the Hawking  spectrum contains at least one
particle species that interacts sufficiently weakly to escape the
neutron star before decaying. Such a state must belong to physics
beyond the Standard Model (SM). Possible candidates include
heavy neutrinos, dark photons, moduli fields, axion-like particles, or
other feebly interacting states ~\cite{Baker:2022rkn}. We
collectively denote such a state by $S$, with mass $m_S$, spin $s_S$,
and $g_S$ internal degrees of freedom.

The Hawking emission rate of $S$ particles with energy between $E$ and
$E+dE$ from a black hole of temperature $T_{\rm BH}$ is
\begin{equation}
\label{eq:Hawkemission}
\frac{d^2N_S}{dt\,dE}
=
\frac{g_S}{2\pi\hbar}
\frac{\mathcal{G}_S(M_{\rm BH},E,s_S)}
{e^{E/T_{\rm BH}}-(-1)^{2s_S}},
\end{equation}
where $\mathcal{G}_S(M_{\rm BH},E,s_S)$ is the absorption coefficient
(graybody factor). This factor arises from gravitational scattering of
particles outside the horizon and must generally be computed
numerically~\cite{Page:1976df}. It can be expressed in terms of the
spin-dependent absorption cross section as
\[
\mathcal{G}_S(M_{\rm BH},E,s_S)
\propto
E^2\,\sigma_S(M_{\rm BH},E),
\]
where $\sigma_S$ accounts for the transmission of the emitted particle
through the curvature potential surrounding the black
hole~\cite{MacGibbon:1990zk}. The overall normalization of the emission
rate is set by
$(2\pi\hbar)^{-1}\simeq2.42\times10^{23}\,
{\rm GeV^{-1}\,s^{-1}}$. The spectrum peaks at
$E_{\rm peak}\simeq1.6\,T_{\rm BH}$, with a small boson--fermion
difference arising from quantum statistics, and reaches a maximum value $\sim 7\,g_S\,10^{21}\,
{\rm GeV^{-1}\,s^{-1}}$.

A massive particle species is efficiently emitted once the black hole
temperature exceeds its rest mass, $T_{\rm BH}\gtrsim m_S$. From
Eq.~(\ref{eq:TMBH}), the condition $T_{\rm BH}\sim m_S$ corresponds to
\begin{equation}
T_{\rm BH}\sim m_S
\quad\Longrightarrow\quad
M_{\rm BH}
\simeq
10^{8}\,{\rm kg}
\left(\frac{100\,{\rm GeV}}{m_S}\right),
\end{equation}
below which $S$ particles are abundantly produced.

If the initial black hole temperature is sufficiently high,
so that all relevant particle species are
relativistic, the emitted Hawking energy is approximately
equipartitioned among the available degrees of freedom. Consequently,
the fraction of the total Hawking energy carried by a given particle
species $S$ is, up to spin-dependent graybody corrections,
\begin{equation}
f_{S|H}
\simeq
\frac{g_S}{g_{\rm H}},
\end{equation}
where $g_S$ is the number of internal degrees of freedom of the
particle $S$, while $g_{\rm H}$ denotes the total number of
relativistic degrees of freedom emitted by the black hole. This fraction changes only if additional particle species become
kinematically accessible.

For the prompt evaporation bursts considered here, it is convenient to
work with the time-integrated Hawking spectrum,
\begin{equation}
\frac{dN_S}{dE}
\equiv
\int_{\rm evap}
dt\,
\frac{d^2N_S}{dt\,dE},
\label{eq:time_int_spectrum}
\end{equation}
where, for the microscopic black holes considered here, the integration
extends over the entire evaporation history. The corresponding energy
spectrum, $E\,dN_S/dE$, peaks at
\[
E_{\rm peak}\simeq5.8\,T_{\rm BH}.
\]

The total number of emitted particles of species $S$ is
\begin{equation}
N_S
=
\int_0^\infty
dE\,
\frac{dN_S}{dE}.
\end{equation}
For sufficiently hot black holes and light particle species,
$m_S\ll T_{\rm BH0}$, one finds
\begin{equation}
N_S
\sim
10^{21\text{--}22}\,g_S
\left(\frac{1~{\rm PeV}}{T_{\rm BH0}}\right)^2,
\label{eq:NSapprox}
\end{equation}
where the numerical coefficient depends on the particle spin through
the graybody factors. 
For
$T_{\rm BH}^{\rm init}=1~{\rm PeV}$,
the total yield lies in the range
$2\times10^{20}$--$10^{22}$
per internal degree of freedom,
depending on the particle spin.

 The corresponding total energy emitted into the species $S$ is
\begin{equation}
E_{S,\rm tot}
=
\int_0^\infty
dE_S\,
E_S\,
\frac{dN_S}{dE_S}
\simeq
f_{S|H}\,M_{\rm BH},
\label{eq:ES_tot}
\end{equation}
where the last relation follows  from energy conservation,
$\sum_iE_{i,\rm tot}\simeq M_{\rm BH}$, together with the approximate
equipartition of the Hawking emitted energy among all relativistic
degrees of freedom.

\subsection{Secondary Spectrum}

We now turn to the observable secondary neutrino spectrum produced by
the decay of the unstable particles $S$, which act as mediators,  emitted in Hawking radiation.
The decay products inherit a substantial fraction of the parent-particle
energy, producing broad secondary spectra whose shape depends on the
decay kinematics.
Throughout this work we focus on the neutrino decay channel, since neutrinos can reach the detector propagating over astrophysical distances essentially without attenuation.
The resulting neutrino spectrum is broad and non-thermal, reflecting
both the decay kinematics of the parent particles and the continuous
increase of the black hole temperature during evaporation. It extends
up to energies of order the maximum Hawking temperature attained during
the burst, with the highest-energy neutrinos emitted during the final
stages of black hole evaporation.

\subsubsection{Secondary spectrum from decays $S\rightarrow\nu+\cdots$}

Let us consider a single unstable escaping particle species $S$ emitted
in Hawking radiation with instantaneous spectrum
$d^2N_S/(dt\,dE_S)$. The instantaneous secondary neutrino spectrum is
obtained by convolving the primary Hawking spectrum with the neutrino
spectrum produced in the decay of a parent particle of energy
$E_S$~\cite{MacGibbon:1990zk},
\begin{align}
\frac{d^2N_\nu^{\rm BH}}{dt\,dE_\nu}
&=
\int_{E_S^{\rm min}}^\infty
dE_S\,
\frac{d^2N_S}{dt\,dE_S}\,
{\rm Br}\,
\frac{dN_{S\rightarrow\nu}(E_S,E_\nu)}
{dE_\nu},
\label{eq:convolnu}
\end{align}
where
${\rm Br}\equiv{\rm Br}(S\rightarrow\nu+\cdots)$
is the branching ratio into neutrinos, and
$dN_{S\rightarrow\nu}/dE_\nu$
is the neutrino spectrum produced in the decay of a single parent
particle of energy $E_S$.
Integrating Eq.~\eqref{eq:convolnu} over the full evaporation history
defines the time-integrated neutrino spectrum from a single black hole
evaporation event,
\begin{equation} \label{eq:NBHnu}
\frac{dN_\nu^{\rm BH}}{dE_\nu}
\equiv
\int_{\rm evap}
dt\,
\frac{d^2N_\nu^{\rm BH}}{dt\,dE_\nu}.
\end{equation}
This quantity represents the total neutrino yield per black hole
evaporation and will be used in the calculation of the diffuse neutrino
flux in Sec.~\ref{sec:ObsSign}.
For the two-body decay
$S\rightarrow\nu\bar\nu$, assuming isotropic decays in the $S$ rest
frame and negligible neutrino masses, the daughter spectrum in the
stellar frame is the familiar box distribution \cite{Ibarra:2012dw},
\begin{equation}
\frac{dN_{S\rightarrow\nu}}{dE_\nu}
=
\frac{2}{E_S\beta_S}\,
\Theta(E_\nu-E_{\nu,\min})
\Theta(E_{\nu,\max}-E_\nu),
\label{eq:boxkernel}
\end{equation}
where
$$
E_{\nu,\min/\max}
=
\frac{E_S}{2}(1\mp\beta_S),
\qquad
\beta_S=\sqrt{1-\frac{m_S^2}{E_S^2}}.
$$
 The distribution is normalized to two neutrinos per decay. The minimum
parent energy required to produce a neutrino of energy $E_\nu$ is
$$
E_S^{\rm min}(E_\nu)
=
\max\!\left[
m_S,\,
E_\nu+\frac{m_S^2}{4E_\nu}
\right].
$$

For non-relativistic parent particles
($E_S\simeq m_S$), the box collapses to a narrow line at
$E_\nu\simeq m_S/2$, whereas for ultra-relativistic parents it becomes
approximately flat over the interval $0<E_\nu<E_S$.

\begin{figure*}[t]
    \centering
    \includegraphics[width=7.8cm,height=4.9cm]{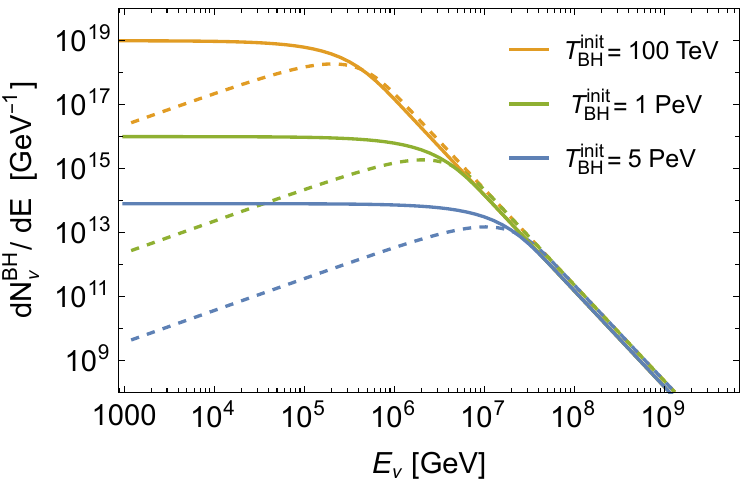}\quad\quad\quad
    \includegraphics[width=7.8cm,height=4.9cm]{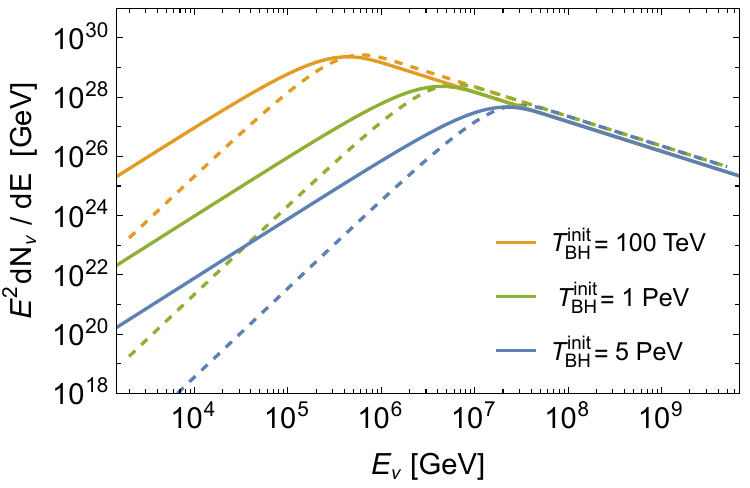}
  \caption{
 {\small {\it Left panel:} Secondary time-integrated spectrum of $\nu$ particles for 
$E_S \gg m_S$. Dashed lines show the corresponding primary spectrum. 
{\it Right panel:} Secondary time-integrated spectrum of $\nu$ particles 
weighted by $E^2$ for $E_S \gg m_S$. Dashed lines show the corresponding 
primary spectrum. }
}
\label{fig:sec_int}
\end{figure*}

For
$m_S\ll T_{\rm BH}$,
convolution of the Hawking spectrum with the two-body decay kernel
produces a broad secondary neutrino spectrum extending up to
$E_\nu\sim T_{\rm BH}$.
Relative to the primary spectrum, the convolution enhances the
low-energy neutrino yield while preserving the asymptotic
$E_\nu^{-3}$ behavior at high energies, up to an
$\mathcal{O}(1)$ normalization factor.
These features are illustrated in
Fig.~\ref{fig:sec_int}.


\section{Observable Signals}\label{sec:ObsSign}

The evaporation of microscopic black holes inside neutron stars may
produce observable neutrino fluxes through the decay of long-lived
particles emitted in Hawking radiation. Depending on the source
distribution, two complementary search strategies can be envisaged.
The first is the detection of an individual nearby neutron star as a
point source. The second is the cumulative emission from the Galactic
neutron star population, which gives rise to an extended neutrino
signal concentrated toward the Galactic Center.

The search for cumulative emission from a population of Hawking
radiators has a long history. Early studies employed the diffuse
$\gamma$-ray background to constrain the cosmological abundance of
evaporating primordial black holes \cite{Page:1976wx,Carr:1976zz},
while more recent works have investigated analogous signatures from
compact objects and dark matter in the Galaxy
\cite{Leane:2021ihh,Bose:2021yhz,   Nguyen:2022zwb,Acevedo:2024ttq, Bose:2024wsh}. Motivated by these
developments, we first consider the neutrino signal from an individual
neutron star before turning to the cumulative Galactic emission.

\subsection{Flux from a Single Neutron Star}

The differential neutrino flux at Earth from a single burst produced by
a neutron star at distance $D$ is
\cite{Leane:2021ihh}
\begin{equation}
\frac{d\Phi_\nu}{dE_\nu}
=
\frac{1}{4\pi D^2}
\int dE_S\;
\frac{dN_S}{dE_S}\;
P_{\rm dec}\;
{\rm Br}\;
\frac{dN_{S \rightarrow \nu}}{dE_\nu}\,,
\label{eq:flux_factorized_burst}
\end{equation}
where $D$ is the distance to the source and
$P_{\rm dec}(E_S,r_1,r_2)$ denotes the probability that $S$ decays between radii $r_1$ and $r_2$, measured from the center of the stellar object,
\begin{equation}
P_{\rm dec}(E,r_1,r_2)
=
\exp\!\left[-\frac{r_1}{\ell_{\rm dec}(E)}\right]
-
\exp\!\left[-\frac{r_2}{\ell_{\rm dec}(E)}\right].
\end{equation}

The observation or non-observation of such bursts constrains the rate of
black hole formation inside neutron stars and therefore the underlying
dark matter capture rate, DM--nucleon interactions, and properties of
the beyond-the-Standard-Model particles responsible for the escaping
radiation.

\subsection{Diffuse Neutrino Flux from the Galactic Center}

The large neutron star population expected in the Galactic Center,
combined with the enhanced dark matter density in this region, makes it
the dominant Galactic source of the cumulative neutrino signal. The
diffuse flux is obtained by integrating the contribution of individual
neutron stars over the Galactic neutron star population, weighted by the
local dark matter capture rate. The calculation therefore requires the Galactic dark matter density profile and the spatial distribution of neutron stars.

The Galactic dark matter density is modeled with a generalized Navarro–Frenk–White (NFW) profile
\cite{Navarro:1995iw,Navarro:1996gj},
\begin{equation}
\rho_{\rm DM}(r) =
\rho_{{\rm DM}\odot}
\frac{(r/r_s)^{-\gamma_{\rm dm}}}{\left(1+r/r_s\right)^{3-\gamma_{\rm dm}}},
\end{equation}
where $r$ denotes the galactocentric radius, $r_s$ is the characteristic scale radius,
and $\gamma_{\rm dm}$ parameterizes the logarithmic inner slope.
The normalization $\rho_{{\rm DM}\odot}$ is fixed by requiring consistency with the
local DM density at the Solar position, $\rho_\odot \simeq 0.3$--$0.4~\mathrm{GeV\,cm^{-3}}$
\cite{Catena:2009mf, deSalas:2019pee}.

The generalized form allows for deviations from the canonical NFW cusp ($\gamma_{\rm dm}=1$),
which may arise due to baryonic effects such as adiabatic contraction or feedback-driven
core formation \cite{Blumenthal:1985qy,Gnedin:2003rj}.
Other commonly used parameterizations include the Einasto and Burkert profiles with $\gamma_{\rm dm} <1$ \cite{Einasto:1965, Navarro:2003ew, Navarro:2008kc}.

Steeper cusps, such as the Moore profile, $\gamma_{\rm dm}=-1.5$, would further enhance the expected signal from
the Galactic Center. The choice of halo profile therefore represents
one of the main astrophysical uncertainties
\cite{Moore:1999gc,Iocco:2011jz}.

\subsubsection{Galactic Neutron Star Population}

The Milky Way is expected to host a total neutron star population of
order $10^{8}$--$10^{9}$, although only a small fraction are directly
observed owing to selection effects
\cite{FaucherGiguere:2006,Lorimer:2006,Keane:2008,Sartore:2009wn}.
For the present work, the most relevant population is that residing in
the Galactic Center, where the enhanced dark matter density leads to the
largest capture rates. Observations of massive O/B stars within the
central parsec indicate ongoing massive-star formation
\cite{Genzel:2003cn,Levin:2003kp} 
strongly supporting the existence of a substantial neutron star
population in this environment.

We adopt the neutron star distribution obtained by
Generozov \emph{et al.}~\cite{Generozov:2018niv}, based on a
Fokker--Planck treatment of compact-object formation and dynamical
evolution in the Galactic Center. Following
Ref.~\cite{Nguyen:2022zwb}, we use the ``Fiducial ($\times10$)''
benchmark, corresponding to an enhanced star-formation history, and
assume a characteristic neutron star mass
$M_{\rm NS}=1.5\,M_\odot$. The resulting number-density profile is
\begin{equation}
\label{eq:NSprofile}
n_{\rm NS}(r)=
\left\{
\begin{aligned}
&5.98\times10^3
\left(\frac{r}{\rm pc}\right)^{-1.7}
{\rm pc}^{-3},
&&0.1<\frac{r}{\rm pc}<2,
\\
&2.08\times10^4
\left(\frac{r}{\rm pc}\right)^{-3.5}
{\rm pc}^{-3},
&&2<\frac{r}{\rm pc}<100.
\end{aligned}
\right.
\end{equation}

Integrating Eq.~\eqref{eq:NSprofile} over the inner
$100\,{\rm pc}$ yields
$
N_{\rm NS}\simeq1.6\times10^6,
$
which we adopt throughout this work when computing the cumulative
Galactic center neutrino emission.

\subsubsection{Integrated Neutrino Flux from the Galactic Center}
\label{sec:DiffuseCalculation}

The cumulative neutrino signal is obtained by summing the contribution
from the entire Galactic-center neutron star population. Since each
black hole evaporation event is triggered by the accumulation and
collapse of a dark matter core of mass $M_{\rm Ch}$, the relevant
quantity is the dark matter capture rate rather than the annihilation
rate.

The total dark matter capture rate within the region
$r_1<r<r_2$ is
\begin{equation}
C_{\chi,\rm tot}^{\rm NS}
=
4\pi
\int_{r_1}^{r_2}
r^2
n_{\rm NS}(r)
C_\chi(r)\,
dr,
\end{equation}
where $n_{\rm NS}(r)$ is the neutron star number density and
$C_\chi(r)$ is the dark matter capture rate of an individual neutron
star at galactocentric radius $r$.

The corresponding capture power is
\begin{equation}
\dot E_{\rm cap}^{\rm tot}
=
m_\chi
C_{\chi,\rm tot}^{\rm NS}
=
4\pi
m_\chi
\int_{r_1}^{r_2}
r^2
n_{\rm NS}(r)
C_\chi(r)\,
dr.
\label{eq:captot}
\end{equation}

\begin{figure}
    \centering
    \includegraphics[width=0.9\linewidth]{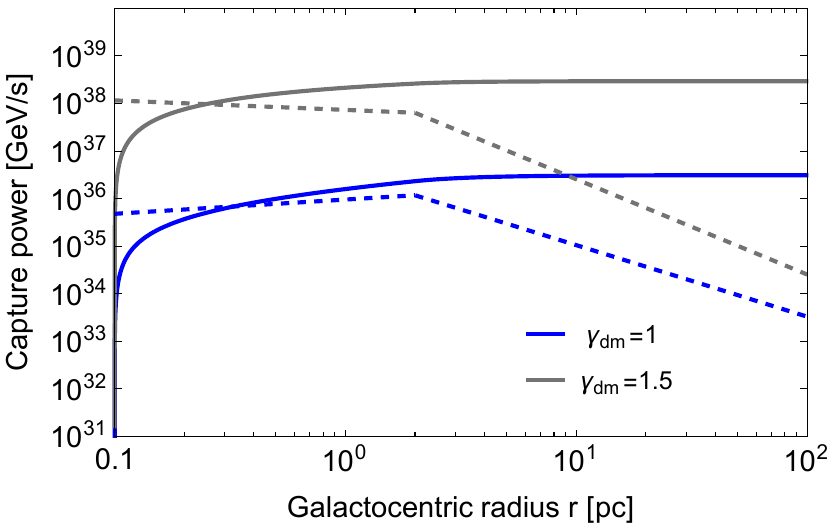}
    \caption{{\small 
    Radial distribution of the dark matter capture power produced by the
Galactic neutron star population. The dashed curves show the differential
contribution to the total capture luminosity per logarithmic radial
interval ${d\dot E_{\rm cap}^{\rm tot}}/{d\ln r}$
while the solid curves show the cumulative capture luminosity, $\dot E_{\rm cap}^{\rm tot}(<r)$.
Results are shown for generalized NFW dark matter profiles with inner
slopes $\gamma_{\rm dm}=1$ (blue) and $\gamma_{\rm dm}=1.5$ (gray). The differential contribution peaks within the central few parsecs,
while the cumulative capture power approaches its asymptotic value
within the inner $\mathcal{O}(100)\,{\rm pc}$ of the Galaxy. 
}}
    \label{fig:CaptPower}
\end{figure}

The strong dependence of Eq.~(\ref{eq:captot}) on the inner dark matter
halo profile follows directly from the radial scaling of the integrand.
Since
$n_{\rm NS}(r)\propto r^{-1.7}$ and, for saturated capture,
$C_\chi(r)\propto\rho_\chi(r)\propto r^{-\gamma_{\rm dm}}$,
one finds $d \dot E_{\rm cap}^{\rm tot}\propto
r^{0.3-\gamma_{\rm dm}}$ for $r<2~{\rm pc}$.
Increasingly cuspy dark matter profiles therefore receive a larger
contribution from neutron stars located closest to the Galactic Center
(Fig.~\ref{fig:CaptPower}).  
The capture power will be transformed to luminosity via BH evaporation. For NFW profile it is 
\begin{equation}
\dot E_{\rm cap}^{\rm tot}
\simeq
3.1\times10^{36}
~{\rm GeV\,s^{-1}}
\simeq
5\times10^{33}
~{\rm erg\,s^{-1}},
\end{equation}
which is comparable to the solar luminosity,
$\dot E_{\rm cap}^{\rm tot}\simeq1.3\,L_\odot$.
Equivalently, the captured power corresponds to the evaporation of a
total microscopic black-hole mass of $\dot M_{\rm BH}^{\rm tot}\simeq5.5\times10^9~{\rm kg\,s^{-1}}$
by the Galactic-center neutron star population. 

The black hole formation (and evaporation) rate per neutron star is
\begin{equation}
\Gamma_{\rm BH}(r)
\simeq
\frac{m_\chi C_\chi(r)}{M_{\rm Ch}},
\end{equation}
since each evaporation event corresponds to the collapse of a dark
matter core of mass $M_{\rm Ch}$. The unresolved neutrino flux from the
Galactic center neutron star population is therefore
\begin{equation}
E^2_\nu\frac{d\Phi_\nu}{dE_\nu}
=
\frac{1}{4\pi}
\int
dV\,
\frac{n_{\rm NS}(r)}
{D(r)^2}
\,
\Gamma_{\rm BH}(r)
\,
E^2_\nu
\frac{dN_\nu^{\rm BH}}{dE_\nu}
\,
P_{\rm dec}
\label{eq:flux_BH_final}
\end{equation}
where $D(r)$ is the distance from the source to the observer,
$P_{\rm dec}$ denotes the probability that the mediator decays outside
the neutron star within the relevant observational region, and
$dN_\nu^{\rm BH}/dE$ is the time-integrated neutrino spectrum produced
by a single black hole evaporation event.

Since the spatial extent of the Galactic center neutron star population
($\lesssim100\,{\rm pc}$) is much smaller than its distance from the
Earth ($D_{\rm GC}\simeq8.3\,{\rm kpc}$), we approximate
$D(r)\simeq D_{\rm GC}$. Equation~(\ref{eq:flux_BH_final}) then reduces
to
\begin{equation}
E^2_\nu\frac{d\Phi_\nu}{dE_\nu}
\simeq
\frac{\dot E_{\rm cap}^{\rm tot}}
{4\pi D_{\rm GC}^2}
\,
\frac{E^2}{M_{\rm Ch}}
\frac{dN_\nu^{\rm BH}}{dE}
\,
P_{\rm dec}.
\label{eq:GCflux}
\end{equation}

Using the neutron star distribution of
Eq.~(\ref{eq:NSprofile}), integrating over
$0.1~{\rm pc}<r<100~{\rm pc}$, and approximating the source distance by
$D(r)\simeq D_{\rm GC}=8.33\pm0.35~{\rm kpc}$, the capture luminosity
received at Earth is
\begin{equation}
\frac{\dot E_{\rm cap}^{\rm tot}}
{4\pi D_{\rm GC}^2}
\simeq
4\times10^{-10}
{\rm GeV\,cm^{-2}\,s^{-1}},
\quad
\gamma_{\rm dm}=1\,.
\end{equation}
Cases with $\gamma_{\rm dm}>1$ should be regarded as optimistic benchmarks. For
example, for $\gamma_{\rm dm}=1.5$ the flux is enhanced by roughly two orders of
magnitude, while shallower or cored halo profiles lead to
correspondingly smaller fluxes. Using $M_{\rm Ch}\simeq f_{S|H}^{-1}E_{S,\rm tot}$
together with the
representative values
\[
f_{S|H}\sim10^{-2},
\qquad
P_{\rm dec}\sim1,
\qquad
\frac{E^2_\nu}{E_{S,\rm tot}}
\frac{dN_\nu^{\rm BH}}{dE_\nu}
\sim0.3,
\]
one obtains the characteristic energy-weighted neutrino flux,   
\begin{equation} \label{eq:power_estimate}
E^2_\nu\frac{d\Phi_\nu}{dE_\nu}
\sim
10^{-12}
{\rm GeV\,cm^{-2}\,s^{-1}}\,.
\end{equation}
Steeper inner dark matter profiles increase the predicted flux,
whereas shallower or cored profiles reduce it. The benchmark flux is
determined by three ingredients: the dark matter capture power, the
properties of the long-lived mediator, and the Hawking-induced neutrino
spectrum.

\subsubsection{ROI Intensity}

The flux estimate in Eq.~(\ref{eq:power_estimate}) corresponds to the
total emission from the neutron star population within the adopted
Galactic-center region. For comparison with Galactic-center searches, it is convenient to
consider the cumulative neutrino flux enclosed within a circular region
of interest (ROI) of angular radius $\theta$,
\begin{equation}
\Phi_\nu(<\theta)
=
\int_{\Delta\Omega(\theta)}
d\Omega\,
\frac{d\Phi_\nu}{d\Omega},
\end{equation}
where
$
\Delta\Omega(\theta)=2\pi(1-\cos\theta)
$
is the corresponding solid angle. We then define the average
energy-weighted neutrino intensity within the ROI as
\begin{equation}
I_\nu(\theta)\equiv
\frac{1}{\Delta\Omega(\theta)}
E_\nu^2
\frac{d\Phi_\nu(<\theta)}{dE_\nu}.
\end{equation}
where $ 
\Delta\Omega(\theta)
\simeq
\pi\theta^2,
$
for $\theta\ll1$.

Assuming spherical symmetry, the corresponding cumulative capture power,
$\dot E_{\rm cap}(<\theta)$, is obtained by integrating the capture power
density $n_{\rm NS}(r)\,m_\chi C_\chi(r)$,
over the volume enclosed by the  radius
 $D_{\rm GC}\tan\theta$.
Table~\ref{tab:ROIflux} summarizes the average energy-weighted neutrino
intensity for representative Galactic-center regions of interest,
providing benchmark predictions for comparison with localized and
template-based neutrino searches.

\begin{table}[t]
\centering
\begin{tabular}{c|c|c}
\hline\hline
$\theta$ &
$\langle I_\nu\rangle$ ($\gamma_{\rm dm}=1$) &
$\langle I_\nu\rangle$ ($\gamma_{\rm dm}=1.5$)
\\
(deg) &
(${\rm GeV\,cm^{-2}\,s^{-1}\,sr^{-1}}$) &
(${\rm GeV\,cm^{-2}\,s^{-1}\,sr^{-1}}$)
\\
\hline
0.2 & $9.7\times10^{-6}$ & $9.2\times10^{-4}$ \\
0.3 & $4.3\times10^{-6}$ & $4.1\times10^{-4}$ \\
0.5 & $1.6\times10^{-6}$ & $1.5\times10^{-4}$ \\
1   & $3.9\times10^{-7}$ & $3.7\times10^{-5}$ \\
2   & $9.8\times10^{-8}$ & $9.2\times10^{-6}$ \\
5   & $1.6\times10^{-8}$ & $1.5\times10^{-6}$ \\
\hline
\end{tabular}
\caption{Average energy-weighted neutrino intensity within a circular
region of interest (ROI) of angular radius $\theta$ centered on the
Galactic Center, for the benchmark NFW profile  and a
cuspy profile. The corresponding integrated flux has
already saturated for $\theta\gtrsim0.1^\circ$.}
\label{tab:ROIflux}
\end{table}

\subsubsection{Expected Event Rates}

During propagation over Galactic distances, neutrino oscillations
average out the flavor composition. Consequently, the neutrino flux
arriving at Earth is expected to be approximately flavor democratic,
$$
\nu_e:\nu_\mu:\nu_\tau\simeq1:1:1,
$$
almost independently of the production mechanism. Each flavor therefore
carries approximately one third of the total neutrino flux. We next estimate the corresponding neutrino number flux and detection rate.

Let us now estimate the  neutrino number flux per logarithmic energy interval.
For illustration, we consider the benchmark energy
$E_\nu\simeq1~{\rm PeV}$. 
This corresponds to the characteristic energy scale of the
time-integrated spectrum from microscopic black holes with initial
Hawking temperature
$
T_{\rm BH}^{\rm init}\sim 200~{\rm TeV},
$
for which the energy-weighted neutrino spectrum peaks at
$E_\nu\sim{\rm PeV}$.
Assuming a standard NFW halo ($\gamma_{\rm dm}=1$), differential neutrino number flux per logarithmic energy interval is
\begin{equation}
\frac{dN_\nu}{dA\,dt\,d\ln E_\nu}
=
\frac{1}{E_\nu}
E^2_\nu\frac{d\Phi_\nu}{dE_\nu}
\sim
10^{-18}
~{\rm cm^{-2}\,s^{-1}},
\end{equation}
where we have used
$E_\nu\simeq1~{\rm PeV}$ and 
Eq.~\eqref{eq:power_estimate}.

For a detector with effective area
$A_{\rm eff}\sim1~{\rm km}^2=10^{10}~{\rm cm^2}$, this corresponds to a
neutrino crossing rate
\begin{equation}
\dot N_\nu
=
A_{\rm eff}
\frac{dN_\nu}{dA\,dt\,d\ln E_\nu}
\sim
10^{-8}
~{\rm s^{-1}},
\end{equation}
or approximately
$ N_\nu\sim0.3~{\rm yr^{-1}}$,
per logarithmic energy interval around the spectral peak at
$E_\nu\simeq1~{\rm PeV}$. 
Since the neutrino spectrum is broad 
the total number of neutrinos crossing the detector is expected to be a
few times larger than the differential estimate above. Consequently, one
expects an integrated crossing rate of
$
N_\nu^{\rm tot}
\sim
\mathcal{O}(1-2)
~{\rm yr^{-1}},
$
for the benchmark NFW profile.

\begin{figure}
    \centering
    \includegraphics[width=1.0\linewidth]{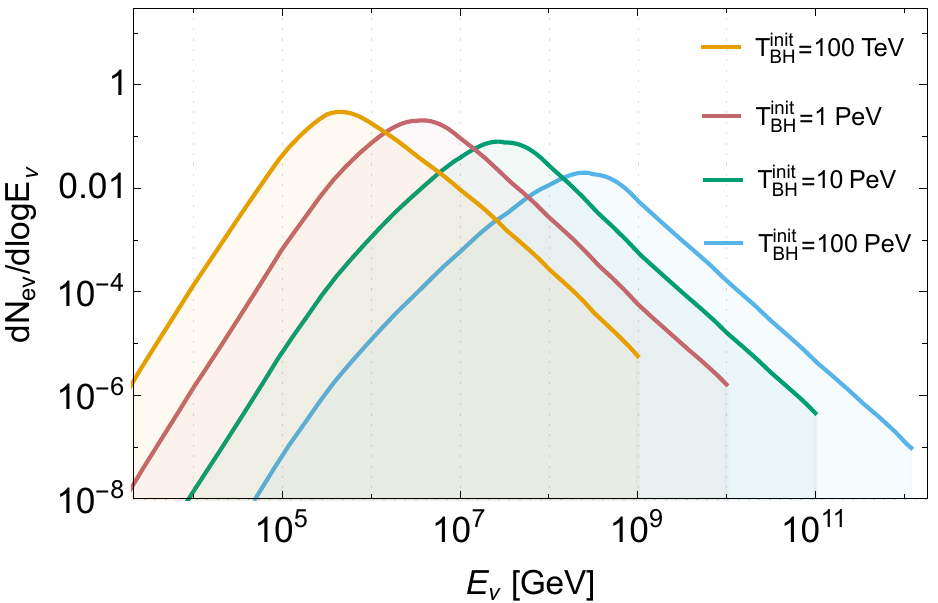}
    \caption{{\small 
 Differential contribution to the expected number of detected neutrino
events per logarithmic energy interval,
for black holes with initial Hawking
temperatures $T_{\rm BH}^{\rm init}=100~{\rm TeV}$,
$1~{\rm PeV}$, $10~{\rm PeV}$ and $100~{\rm PeV}$.
The curves correspond to the neutrino flux folded with the IceCube
effective area $A_{\rm eff}$ for through-going $\nu_\mu$ neutrinos, assuming a 10-year
observation time and cuspy 
dark matter halo profile. The area under each curve gives the total expected number of detected
events.
}}
    \label{fig:DetectEvent}
\end{figure}

Only a fraction of these neutrinos produce detectable events. As an illustrative estimate, the probability for a PeV neutrino to
interact while traversing a length $L\sim1~{\rm km}$ of ice with nuclei density $n_N$ is
$$
P_{\rm int}
\simeq
n_N\sigma_{\nu N}L
\sim
10^{-5}\text{--}10^{-4}.
$$
This should be regarded as an interaction probability within the
instrumented volume, not as the full detector efficiency.
A realistic prediction, however, requires folding the neutrino flux with
the published, energy- and direction-dependent IceCube effective area,
\begin{equation} \label{eq:Nevent}
N_{\rm ev}
=
T
\int dE_\nu\,
A_{\rm eff}(E_\nu,\delta_{\rm GC})
\frac{d\Phi_\nu^{\rm GC}}{dE_\nu},
\end{equation}
where $A_{\rm eff}(E_\nu,\delta_{\rm GC})$ is the detector effective area at the declination of
the Galactic Center, 
 $T$ denotes the detector lifetime 
\cite{IceCube:2023ame_Galactic, IceCube-Gen2:2020qha_Rev, IceCube:2020nig_Aeff}. 
The effective area incorporates the neutrino interaction probability,
detector geometry, event selection and reconstruction efficiencies.
Since both the neutrino--nucleon cross section and the effective area
increase with energy, the highest energy part of the spectrum
contributes disproportionately to the detected event rate. The corresponding detector-weighted spectra for a cuspy dark matter profile ($\gamma_{\rm dm}=1.5$) are shown in Fig. ~\ref{fig:DetectEvent}.

For favorable microscopic parameters, an NFW dark matter halo profile,
and a 10-year IceCube exposure, we predict
$\mathcal{O}(10^{-2})$ detected events for
$T_{\rm BH}^{\rm init}=100~{\rm TeV}$. Comparable rates are obtained
for $T_{\rm BH}^{\rm init}=1~{\rm PeV}$, while the event rate
decreases at higher Hawking temperatures. A cuspy halo with
$\gamma_{\rm dm}=1.5$ enhances the signal by approximately two orders
of magnitude, yielding
$N_{\rm ev}\sim1$ for
$T_{\rm BH}^{\rm init}=100~{\rm TeV}$--$1~{\rm PeV}$, and
$\mathcal{O}(10^{-1})$ for
$T_{\rm BH}^{\rm init}=10~{\rm PeV}$. Since the Galactic component
constitutes only about $10\%$ of the observed high-energy
astrophysical neutrino flux, the predicted signal may contribute at
the $\mathcal{O}(1$--$10)\%$ level to the Galactic neutrino flux under
favorable astrophysical conditions.

The predicted event rate remains modest even for optimistic dark matter
profiles, primarily because the available power is limited by the dark
matter capture rate and only a small fraction of the emitted neutrinos
are detected. Nevertheless, next-generation facilities such as
IceCube-Gen2 \cite{IceCube-Gen2:2020qha}, KM3NeT \cite{KM3Net:2016zxf}  GRAND \cite{GRAND:2018iaj} and RNO-G \cite{RNO-G:2020rmc}  may significantly improve the
prospects for detection. The predicted signal is concentrated in the
$10~{\rm TeV}$--EeV energy range, where the atmospheric neutrino
background is strongly suppressed, making Galactic-center searches
particularly promising.

\subsection{Extragalactic Diffuse Neutrino Flux}

The mechanism discussed above is expected to operate not only in the
Milky Way but also in neutron stars throughout the Universe.
Consequently, it naturally predicts a diffuse extragalactic neutrino
background. As a simple benchmark, we estimate its magnitude under the
assumption that the Milky Way is representative of an average galaxy.

Assuming a comoving number density of Milky Way-like galaxies,
\begin{equation}
\label{eq:ng}
n_g
\simeq
10^{-2}\ {\rm Mpc^{-3}},
\end{equation}
the corresponding cosmological neutrino emissivity is obtained by
multiplying the Galactic neutrino luminosity by the galaxy number
density. The diffuse extragalactic neutrino intensity is then
approximately
\begin{equation}
E_\nu^2\frac{d\Phi_\nu^{\rm EG}}{dE_\nu\,d\Omega}
\simeq
\frac{c}{4\pi H_0}\,
\xi_z\, n_g
E_\nu^2\frac{dL_\nu^{\rm GC}}{dE_\nu}
\label{eq:EGestimate}
\end{equation}
where $\xi_z\simeq2$--3 parametrizes the enhancement due to the
cosmological evolution of the source population
\cite{Waxman:1998yy,Murase:2016gly_WaxEGnu}. 
The energy-weighted
Galactic neutrino luminosity defined above is
\begin{equation}
E_\nu^2\frac{dL_\nu^{\rm GC}}{dE_\nu}
=
4\pi D_{\rm GC}^2
E_\nu^2\frac{d\Phi_\nu^{\rm GC}}{dE_\nu}
\sim
8\times10^{33}\,
{\rm GeV\,s^{-1}},
\end{equation}
where we used the benchmark Galactic-center neutrino flux obtained in
Eq.~\eqref{eq:power_estimate}. 
This benchmark should be regarded
as conservative, since it includes only the neutron star population
within the central $100\,{\rm pc}$ and neglects contributions from the
Galactic bulge and disk.
The corresponding cosmological neutrino emissivity is
\begin{equation}
n_g
E_\nu^2\frac{dL_\nu^{\rm GC}}{dE_\nu}
\sim
3\times10^{-42}
~
{\rm GeV\,cm^{-3}\,s^{-1}}.
\end{equation}
Using Eq.~\eqref{eq:EGestimate} together with
${c}/{H_0}\simeq1.3\times10^{28}\,{\rm cm}$ yields the isotropic
diffuse neutrino intensity per unit solid angle,
\begin{equation}
E_\nu^2\frac{d\Phi_\nu^{\rm EG}}{dE_\nu\,d\Omega}
\sim
(6-9)\times10^{-15}
~
{\rm GeV\,cm^{-2}\,s^{-1}\,sr^{-1}},
\end{equation}
where the range reflects the uncertainty in the source-evolution factor
$\xi_z$.

Integrating over the full sky, the diffuse
extragalactic neutrino flux for a standard NFW halo
($\gamma_{\rm dm}=1$) is $4\pi$ times larger.
Steeper inner dark matter profiles can enhance this estimate by roughly
two orders of magnitude, reaching
$E_\nu^2d\Phi_\nu^{\rm EG}/dE_\nu\sim10^{-11}\,
{\rm GeV\,cm^{-2}\,s^{-1}}$.

The predicted extragalactic contribution is not shown in
Fig.~\ref{fig:FlucIceCube}, but it would exhibit essentially the same
spectral shape as the Galactic-center signal, differing only in its
overall normalization. Under the benchmark assumptions adopted here,
the resulting diffuse extragalactic component remains subdominant,
although it may constitute a non-negligible fraction of the observed
high-energy cosmic neutrino background. A larger contribution could
arise in galaxies hosting denser nuclear star clusters, larger neutron
star populations, or enhanced central dark matter densities.

\subsection{Comparison with Diffuse Neutrino Backgrounds}

Current IceCube observations have established a diffuse Galactic
neutrino component concentrated along the Galactic plane, accounting for
roughly $10\%$ of the total astrophysical neutrino flux
\cite{IceCube:2023ame_Galactic}. 
 Its properties are broadly consistent with
models in which neutrinos are produced by interactions of Galactic
cosmic rays with the interstellar medium. 
\cite{Gaggero:2015xza_CRinterpretation}. 
The neutron star population considered
here provides an additional Galactic component concentrated toward the
Galactic Center, which can be constrained through existing Galactic
plane and Galactic Center template analyses. It is therefore useful to
compare the predicted Galactic and extragalactic contributions with the
established astrophysical neutrino backgrounds.

The neutron star population considered in the present work may provide
an additional source of high energy Galactic neutrinos, particularly
towards the Galactic Center where both the neutron star abundance and
the dark matter density are expected to be largest. More
conservatively, the predicted signal can be constrained through
existing Galactic plane and Galactic Center template analyses.

It is therefore useful to compare the predicted Galactic and
extragalactic contributions with the currently established diffuse
high-energy neutrino backgrounds. The dominant component is the nearly
isotropic diffuse extragalactic neutrino flux, commonly attributed to
the superposition of unresolved astrophysical sources; see, e.g.,
Refs.~\cite{Murase:2022dog_EGnu,Fang:2023azx}.

The neutrino signal predicted in this work should be regarded as an
additional contribution to the established astrophysical neutrino
backgrounds. Schematically, the total observed flux may be written as
\begin{equation}
\Phi_\nu^{\rm obs}
=
\Phi_\nu^{\rm EG,astro}
+
\Phi_\nu^{\rm Gal,astro}
+
\Phi_\nu^{\rm EG,NS}
+
\Phi_\nu^{\rm MW,NS},
\label{eq:total_flux}
\end{equation}
where the first two terms denote the conventional extragalactic and
Galactic astrophysical components, while the last two correspond to the
extragalactic and Milky Way contributions from dark matter accumulated
inside neutron stars.

\begin{figure}
    \centering
    \includegraphics[width=1.0\linewidth]{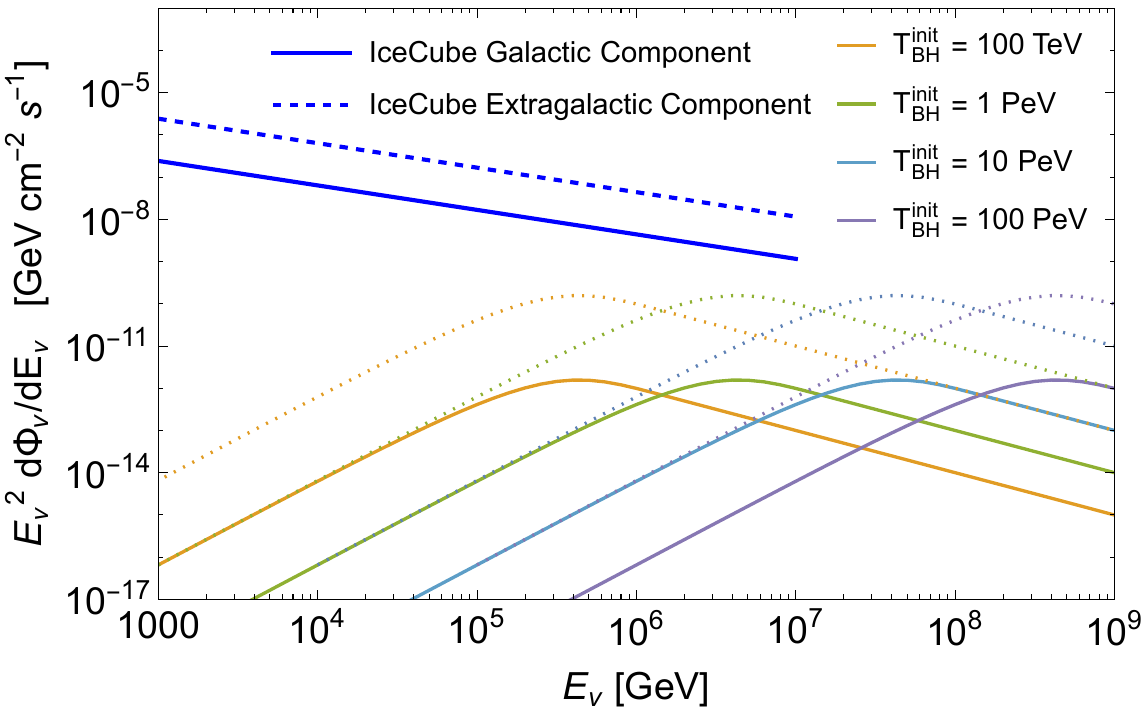}
    \caption{{\small 
Galactic energy-weighted neutrino flux,
$E_\nu^2\,d\Phi_\nu/dE_\nu$, predicted in the present scenario for
benchmark initial black hole Hawking temperatures
$T_{\rm BH}^{\rm init}=100~{\rm TeV}$, $1~{\rm PeV}$,
$10~{\rm PeV}$, and $100~{\rm PeV}$, assuming an NFW dark matter
profile ($\gamma_{\rm dm}=1$, solid curves) and a cuspy profile
($\gamma_{\rm dm}=1.5$, dotted curves). The predictions are compared
with the IceCube best-fit Galactic and diffuse astrophysical
(predominantly extragalactic) neutrino components. The flux normalization is determined by the dark matter capture power
of the Galactic-center neutron star population, while the spectral peak
is set primarily by the initial Hawking temperature.
}}
    \label{fig:FlucIceCube}
\end{figure}

The diffuse astrophysical neutrino flux measured by IceCube is well
described by a power-law spectrum,
$d\Phi_\nu/dE\propto E^{-2.6}$, while the Galactic component exhibits a
similar or slightly softer spectrum and dominates primarily below a few
tens of TeV
\cite{IceCube:2023ame_Galactic,Gaggero:2015xza_CRinterpretation}. By
contrast, the signal predicted in this work exhibits a broad spectral
peak whose position is determined mainly by the initial Hawking
temperature of the evaporating black hole and the mediator decay
kinematics. At energies well above the peak, the secondary neutrino
spectrum approaches the universal Hawking tail,
\begin{equation}  \label{eq:Hawktail}
E^2_\nu\frac{dN_\nu^{\rm BH}}{dE_\nu}
\propto
E^{-1},
\end{equation}
up to graybody and spin-dependent corrections. The scenario of black hole evaporation within a neutron star 
produces a peak that is expected to lie in the
$\gtrsim10~{\rm TeV}$ energy range.

In addition to its spectral signature, the present mechanism exhibits a
distinct spatial morphology. Unlike the conventional Galactic diffuse
emission, which traces the distribution of interstellar gas and cosmic
rays, the neutrino emissivity approximately follows
\begin{equation}
q_\nu(r)
\propto
n_{\rm NS}(r)\,
\rho_{\chi}(r),
\end{equation}
and is therefore expected to be significantly more concentrated toward
the Galactic Center, as illustrated in
Fig.~\ref{fig:CapR}. The corresponding average energy-weighted
neutrino intensities for different Galactic-center regions of interest
are summarized in Table~\ref{tab:ROIflux}. These predictions can be directly compared with localized and template-based Galactic-center searches. 

IceCube has performed dedicated searches for extended Galactic-center
emission with angular extensions between
$0.5^\circ$ and $2^\circ$
\cite{IceCube:2023ujd_ROI}. While current data remain statistically
limited and do not yet place stringent constraints on the present
scenario, this constitutes a well-motivated target for future Galactic-center
searches.

Figure~\ref{fig:FlucIceCube} compares the predicted
energy-weighted neutrino spectrum with the observed diffuse
astrophysical and Galactic neutrino fluxes. In contrast to the
approximately power-law astrophysical backgrounds, the Hawking-induced
signal exhibits a broad spectral peak, preceded by the low-energy rise
 and followed by the characteristic
high-energy behavior given by Eq. \eqref{eq:Hawktail}.
The combination of this distinctive spectral shape and the strong
concentration of the emission toward the Galactic Center provides two
complementary observables for distinguishing the present mechanism
from conventional astrophysical neutrino sources.

The same mechanism naturally contributes to the diffuse extragalactic
neutrino background. If the Milky Way
is representative of an average galaxy, this contribution remains
subdominant. The overall normalization, however, is expected to vary
significantly from galaxy to galaxy owing to differences in neutron
star populations and central dark matter densities
\cite{Neumayer:2020gno}. By contrast, the spectral shape is expected to
be largely universal, being determined primarily by Hawking evaporation
and mediator decays.

\section{Discussion and Conclusions}

In this work, motivated by the emergence of high-energy neutrino astronomy as a powerful probe of new physics, we have investigated high-energy neutrino production from microscopic black holes formed through the gravitational collapse of asymmetric dark matter accumulated inside neutron stars. When the
initial black hole mass lies below the critical value at which
accretion overtakes Hawking evaporation, the newly formed black hole
evaporates rapidly, converting the accumulated dark matter energy into
Hawking radiation. Assuming the existence of long-lived particles
beyond the Standard Model that are emitted in Hawking radiation,
escape the neutron star, and subsequently decay into neutrinos, the
mechanism naturally generates high-energy neutrino emission extending
from the TeV to the EeV regime. While Hawking evaporation of
primordial black holes has long been exploited to constrain their
abundance and, indirectly, the primordial curvature power spectrum on
small scales see e.g. \cite{Carr:2020gox, Dalianis:2018ymb}, the present work
explores a complementary realization in which microscopic black holes
are continuously produced through dark matter collapse inside neutron
stars. Black holes are born hot, with Hawking temperatures above the TeV scale, opening a new avenue for probing both Hawking radiation and the particle nature of dark matter.

A central result of the present work is the long-term evolution of the
captured dark matter cloud, which naturally gives rise to repeated
dark matter collapse--black hole evaporation cycles.  We have shown that the competition
between the dark matter thermalization time and the interval between
successive collapse events leads to two qualitatively distinct
regimes. In the fully thermalized regime, the cloud cools back to the
neutron star temperature before each collapse. By contrast, when the
thermalization time exceeds the collapse cycle, the energy deposited
by successive black hole evaporation events cannot be efficiently
dissipated. The dark matter cloud therefore evolves toward a
quasi-stationary, partially thermalized configuration whose
equilibrium temperature can exceed the neutron star core temperature
by several orders of magnitude. The resulting thermal pressure
supports a hot, spatially extended dark matter cloud, thereby
modifying the initial conditions for subsequent gravitational
collapse while preserving the cyclic formation and evaporation of
microscopic black holes.

Within this framework we computed the Hawking emission spectra,
and the
secondary neutrino spectra arising from mediator decays, and the
resulting Galactic and diffuse extragalactic neutrino fluxes. The
mechanism predicts two distinctive observational signatures. First,
the neutrino spectrum exhibits a broad, non-power-law peak whose
position is determined primarily by the initial Hawking temperature,
followed by the characteristic high-energy Hawking tail. Second, the
Galactic emission is expected to appear as an extended excess centered
on the Galactic Center, approximately tracing the product of the
neutron star distribution and the ambient dark matter density. Unlike
the nearly isotropic diffuse astrophysical neutrino background, this
characteristic morphology provides a powerful discriminator of the
present scenario.

For benchmark Galactic models we find that the expected number of
detected IceCube events is generally small. For an NFW dark matter
profile, the predicted event rate is typically of order
$10^{-2}$ over a ten-year observation period, even under optimistic
microscopic assumptions, including efficient Hawking emission of the
mediator, favorable branching fractions and decay lengths, and
geometrically saturated dark matter capture. Cuspy dark matter distributions, however, enhance the signal
by approximately two orders of magnitude. The relatively low event rate
reflects the finite dark matter capture power together with the limited
neutrino yield per evaporation event. Nevertheless, the current IceCube sample contains of order $10^2$
neutrino events above $100\,{\rm TeV}$. Since the Galactic component
accounts for only about $10\%$ of the observed astrophysical neutrino
flux, the mechanism discussed here may contribute at the
$\mathcal{O}(1$--$10)\%$ level to the Galactic high-energy neutrino
flux for favorable microscopic and astrophysical parameters. The most
promising targets for future searches include the Galactic Center, the
Galactic plane, and nearby dwarf spheroidal galaxies, where enhanced
dark matter densities and favorable astrophysical conditions maximize
the prospects for detection.

The same mechanism also contributes to the diffuse extragalactic
neutrino background. Assuming the Milky Way is representative of a
typical galaxy, this contribution is subdominant. However, galaxies
with denser nuclear star clusters and higher central dark matter
densities may be significantly brighter while retaining the
characteristic spectral shape predicted here.

The present analysis has intentionally remained largely
model-independent. We considered a generic long-lived feebly interacting particle whose
decays produce high-energy neutrinos outside the neutron star.
Many well-motivated extensions of the Standard Model predict
particles with similar properties. The formalism developed here can
therefore be readily applied to a broad class of particle physics
scenarios by specifying the corresponding Hawking branching fractions,
decay channels, and decay lengths. Moreover, the observation of such a
signal would provide strong evidence for asymmetric dark matter. In the
minimal scenarios considered here, microscopic black hole formation
naturally favors heavy or ultra-heavy dark matter candidates in the
absence of self-interactions or in the presence of repulsive
self-interactions. Attractive self-interactions, on the other hand, can substantially
lower the collapse threshold, allowing microscopic black holes to form
for much lighter dark matter particles and thereby extending the
mechanism to a wider range of dark matter masses. The scenario
therefore establishes a direct connection between terrestrial searches
for heavy dark matter and the astrophysical signatures discussed here.

Future neutrino observatories with substantially larger effective
areas, including IceCube-Gen2, KM3NeT, GRAND, and RNO-G
\cite{IceCube-Gen2:2020qha,KM3Net:2016zxf,GRAND:2018iaj},
will significantly improve the sensitivity to the TeV--EeV energy
range explored in this work, with dedicated searches toward the
Galactic Center and other nearby dark matter-rich environments
providing particularly promising targets. Complementary tests may also
arise from the thermal emission of old neutron stars. Future infrared
facilities, including the James Webb Space Telescope (JWST), the
Extremely Large Telescope (ELT), and the Thirty Meter Telescope (TMT),
are expected to be sensitive to neutron-star surface temperatures of
order $T_{\rm NS}\sim10^3\,{\rm K}$ \cite{Raj:2024kjq}, offering an
independent probe of the heating associated with repeated black hole
evaporation. Together, these multi-messenger observations offer encouraging
prospects for testing the proposed mechanism.

Overall, the scenario explored here opens a new observational window onto Hawking evaporation through microscopic black holes  repeatedly produced by dark matter collapse inside neutron stars. It establishes a
direct connection between dark matter physics, black hole
thermodynamics, compact objects, and high-energy neutrino astronomy,
providing a well-defined multi-messenger target for current and future
observations.

\section*{Acknowledgements}
I would like to thank Chris Kouvaris for numerous valuable discussions and comments.


\newpage

\bibliographystyle{apsrev4-2}
\bibliography{references}





\end{document}